\documentclass[a4paper,11pt]{article}
\usepackage{jheppub} 

\usepackage{bm,amsmath,amssymb,slashed,graphicx,%
            enumerate,alltt,xspace,multirow,xcolor,mathrsfs}
\usepackage{fancyvrb}
\usepackage{cancel}

\usepackage{subcaption}

\usepackage{changes}

\definechangesauthor[name=pn, color=red]{pn}

\RequirePackage{braket}
\RequirePackage{graphicx}

\RequirePackage[numbers,sort&compress]{natbib}

\makeatletter
\g@addto@macro\bfseries{\boldmath}
\makeatother

\definecolor{labelkey}{rgb}{0,0.5,0.0}

\usepackage{listings}
\definecolor{darkgreen}{rgb}{0,0.4,0}

\newcommand\mathd{\mathrm{d}}

\newcommand{\as}{\alpha_s}

\newcommand{\mtop}{m}

\newcommand{\POWHEG}{{\tt POWHEG}}
\newcommand{\POWHEGBOX}{{\tt POWHEG BOX}}
\newcommand{\POWHEGRES}{{\tt POWHEG-BOX-RES}}
\newcommand{\POWHEGhvq}{{\tt POWHEG-hvq}}
\newcommand{\hvq}{\POWHEGhvq}
\newcommand{\bbfourl}{{\tt POWHEG-BOX-RES/b\_bbar\_4l}}
\newcommand{\bbfourlshort}{{\tt b\_bbar\_4l}}

\newcommand{\ttMiNNLO}{{\tt MiNNLO-ttbar}}

\newcommand\mtt{m_{t{\bar t}}}
\newcommand\Ecut{E_{\rm cut}}

\newcommand{\thro}{{\tt thr1}}
\newcommand{\thrt}{{\tt thr2}}
\newcommand{\bbfl}{{\tt bb4l}}
\newcommand{\nrr}{NRR}
\newcommand{\mcut}{M_{\rm cut}}

\title{Top-Antitop Production and Decay at Threshold\\ at the LHC
  in QCD Perturbation Theory.}

\preprint{
  \begin{flushright}
    LAPTH-043/26   
  \end{flushright}
}

\author[b]{Paolo Nason,}
\author[a,b]{Giovanni Pelliccioli}
\author[a,b,1]{Emanuele Re\note{On leave of absence from \emph{Laboratoire d'Annecy de Physique Th\'eorique (LAPTh), CNRS, USMB, F-74940 Annecy, France.}}}
\author[a,b]{Luca Rottoli}

\emailAdd{paolo.nason@mib.infn.it}
\emailAdd{giovanni.pelliccioli@unimib.it}
\emailAdd{emanuele.re@mib.infn.it}
\emailAdd{luca.rottoli@unimib.it}

\affiliation[a]{Dipartimento di Fisica G. Occhialini, Universit\`a degli Studi di Milano-Bicocca}
\affiliation[b]{INFN, Sezione di
  Milano-Bicocca, Piazza della Scienza 3,20126 Milano, Italy}

\date{Received: date / Accepted: \today}

\abstract{In this work we consider the production of a top-antitop
  pair at the LHC when the mass of the pair is relatively near to the
  nominal threshold, that is to say to twice the top pole mass.  In
  this regime, enhanced perturbative corrections arise
  that can be computed to all orders in perturbation theory. We present
  three generators of the NLO+PS kind (Next-to-Leading-Order that can
  be interfaced to parton showers) that include these threshold
  enhanced effects.  Using these generators we address the following
  questions: what is the size of enhanced non-relativistic effects
  that are not already present in the well known NLO and NNLO
  perturbative results; what is the size of the contribution from
  these effects that can be loosely attributed to toponium production;
  and to what extent the finite width of the top quark affects threshold
  enhanced corrections. Our generators are relevant for the recent observation
  of enhanced $t{\bar t}$ production near threshold in the pseudoscalar
  channel by the ATLAS and CMS collaborations.}

\keywords{Perturbative QCD, QCD Phenomenology, top quark pair production, spin correlations}

\begin{document}

\maketitle


\section{Introduction}
\label{sec:intro}
In two recent publications the CMS~\cite{CMS:2025kzt} and
ATLAS~\cite{ATLAS:2026dbe} experiments have claimed the observation of
an excess in the $t{\bar t}$ cross section in the threshold region, in
the pseudoscalar spin state. The excess has been loosely interpreted
as due to the effect of a $t{\bar t}$ bound state in the pseudoscalar
channel (dubbed $\eta_t$), and it is often referred to as the
observation of a toponium state at the LHC. Both CMS and ATLAS qualify
this excess as as a cross section for the $\eta_t$ production, equal
to $8.8^{+1.2}_{-1.4}$~pb and $9.3^{+1.4}_{-1.3}$~pb in the CMS and
ATLAS setups, respectively.\footnote{For completeness we recall that
such excesses are partially dependent on the predictions used to
extract them. For instance, in ref.~\cite{ATLAS:2026dbe}, the excess
is also quoted to be $8.5^{+1.2}_{-1.1}$~pb if a different Monte Carlo
is used as a ``baseline'' for the perturbative QCD prediction, and
$13.1^{+1.9}_{-1.7}$~pb if a simplified model is used to model the
formation of a quasi-bound state.}
According to the ATLAS paper, as stated in the abstract,
the excess is consistent with the
formation of a $t{\bar t}$ bound state as predicted by QCD.
CMS makes a similar claim, adding that the excess is above
the perturbative QCD prediction.

In the ATLAS analysis, an attempt is made to fit the excess using
approaches relying upon the non-relativistic limit of perturbative
QCD~\cite{Fadin:1987wz,Fadin:1990wx}, and in particular from
ref.~\cite{Fuks:2024yjj}.  According to the latter approach, the
toponium contribution is found to be $6.43$ pb,\footnote{This cross
section was obtained in ref.~\cite{Fuks:2021xje}, interpolating
previous results from ref.~\cite{Sumino:2010bv}, where the top mass
was set equal to $173$ GeV.} and thus roughly compatible with the
observed effect.

There are a number of questions that need clarification on the issue discussed above. First of all,
a $t{\bar t}$ state produced near threshold in gluon fusion is known to have a prevalent pseudoscalar spin
component. In fact, near threshold the $s$-wave state prevails, and the angular momentum is only given by
the spin. It turns out that the Landau-Yang theorem \cite{Landau:1948kw,Yang:1950rg},
which forbids the coupling of a spin one state to two on-shell photons,
also applies to the pairs of incoming gluons in leading order QCD, so that the prevalent
spin combination is the singlet one, corresponding to a pseudoscalar state. This holds irrespective
of the fact that the $t{\bar t}$ pair is in a bound state or not. This observation is important, because other
enhancement mechanisms are present in the $t{\bar t}$ production near threshold, and they cannot be
qualified as toponium production.

The second point to clarify has to do with the fact that the toponium state does not really exist
as an isolated resonance. This is a well known fact,
and it follows from simple considerations on the non-relativistic dynamics of the $t{\bar t}$ state.
The binding is caused by the coulombic potential, that is of order $\as/r$, $r$ being the size of the
system. By the uncertainty principle, the relative momentum of the $t{\bar t}$ pair must be of order
$p\approx 1/r$. By the virial theorem, the kinetic energy and the potential energy should be comparable in a bound
state so that
\begin{equation}
  \frac{p^2}{\mtop}\approx \frac{\as}{r},
\end{equation}
(where $\mtop$ is the mass of the top quark)
which leads to $r\approx 1/(\as \mtop)$ which in turn leads to $p\approx \as \mtop$, so that we conclude
that the relative velocity $v$ is of order $\as$. The time needed for the top quarks to complete a full orbit
is thus of order $r/v\approx 1/(\as^2 \mtop)$, that is also of the order of the inverse binding energy, as
required by the uncertainty principle. This time is also very close to the system lifetime, that is one
half of the top lifetime. Thus we cannot expect to see a narrow peak due to the toponium.
However, a small bump in the cross section in the threshold region should be visible. In that region the
cross section is rapidly varying, since it overlaps with the opening of the phase space, that is also
smeared by the finite width of the top, so that the area spanned by the peak cannot be separated
unambiguously from the rest of the cross section. On the other hand, from a theoretical point of view,
we can compute the cross section contribution due to the resonances, assuming a fictitiously small top width.
The integral of that cross section is independent of the width for small width,
so we can call that cross section ``toponium cross section'', even if we know that it will be smeared out
by the finite top width.\footnote{These facts are all well known. We have preferred to
  reiterate them here since unfortunately they are
  not always represented correctly in current literature~\cite{Fuks:2024yjj}.}

As a third point, we remind the reader that toponium production is
part of the perturbative QCD cross section. Claiming that there is an
excess above perturbative QCD prediction that is consistent with
toponium production as predicted by QCD is a contradictory statement.
This fact is well known, it has a long history in the
literature~\cite{Braun:1968njz,Melnikov:2014lwa,Beneke:2016jpx} dating
back to similar issues in positronium production, and it was very
recently reiterated in ref.~\cite{Nason:2025hix} (which we will dub NRR
in the following) for the problem at hand.  Thus, the statement
sometimes found in the literature~\cite{Fuks:2024yjj} that the bound
state contribution cannot be obtained in perturbation theory is
incorrect.

As a fourth point, we remind here that the toponium cross section has
always been estimated to be quite small~\cite{Fadin:1987wz,Fadin:1990wx,%
  Beneke:2016jpx}, much less
than the value fitted by CMS and ATLAS using a pseudoscalar model.
Part of the problem is related to the fact that the result of the
calculation of ref.~\cite{Fuks:2024yjj} (dubbed GFRW,
for Green's function reweighting method) is referred to as the toponium
contribution to the top production cross section. What is implemented
there is instead the resummation of perturbative
contributions that are enhanced near threshold,
following a method first proposed in
refs.~\cite{Fadin:1987wz,Fadin:1990wx}, according to ref.~\cite{Sumino:2010bv}.  The full
calculation of the $t{\bar t}$ hadroproduction cross section is known
up to relative order $\as$ (NLO) and $\as^2$ (NNLO), and (of course)
it does include the enhanced contributions that are obtained in GFRW
up to relative order $\as$ and $\as^2$ respectively.

The fifth problem is related to the way data is presented.
The experiments do not provide an unfolded
cross section for top pair production in invariant mass bins.
What they do instead is to fit the data to
an event sample obtained by adding together events obtained with standard
simulation tools, and events obtained by either using the GFRW model, or a
simplified model for the production of an $\eta_t$~\cite{Maltoni:2024tul}
with a cross section normalization that at the end is fitted to the
data. The number quoted by the experiments is typically the cross
section normalization of this extra sample. When using the GFRW model,
ATLAS is in fact attempting to model the full production region by
adding enhanced non-relativistic effects to all orders in perturbation
theory. These effects are included only when very close to threshold,
by requiring that, at the generator level, the invariant mass of the pair
$\mtt$ is less than 350~GeV and the
relative top momentum in the $t{\bar t}$ rest frame ($p^\star$) is less than
50~GeV, a region that we will call the ``ATLAS bin'' in the following.
The combination of the two samples within that bin is however a
delicate issue because of potential double-counting problems, given
that the GFRW contribution should also contain some higher-order
corrections already present in the baseline generator.

The present work is an extension of the work of \nrr{}, where
we consider three ways for computing the threshold-enhanced contributions
to the $t{\bar t}$ cross section at the LHC.\footnote{The relevant
  codes have been made publicly available, as documented in the Appendix.}

The first calculation, that will be referred to as \thro{}, is an
extension of the one presented in \nrr{} that includes threshold
corrections of order higher than those considered there, obtained
using the \hvq{} generator~\cite{Frixione:2007nw}, supplemented
with a calculation of the threshold enhanced corrections\footnote{In
  the present work we will always prefer to refer to ``threshold
  enhanced corrections'' rather than ``non-relativistic
  effects''. It is quite obvious that at the threshold the
  non-relativistic approximation is applicable, but for some of the
  corrections (namely the NLO and NNLO ones) this is not necessary.}
in the narrow width limit. While in \nrr{} 
threshold corrections of order up to $(\as/v)^3$ were considered (where $v$ is
the velocity of the top quarks in the $t{\bar t}$ rest frame), here we
consider corrections up to $(\as/v)^6$, and also the inclusion of all
corrections of even order, and an estimate of the missing corrections
of odd order. In this calculation, finite width effects are included
by suitably modifying the on-shell partonic events generated at the Les-Houches
level~\cite{Alwall:2006yp} (i.e. the partonic events to be fed to the parton-shower generator)
using the mechanism employed in \hvq{}, which is an
implementation of the method introduced in
ref.~\cite{Frixione:2007zp}.

The second calculation, that will be referred to as \thrt{}, deals with the resummation of threshold
effects taking into account the finite width of the top quark.
In this case we started again from the \hvq{} code, but modified the phase-space generation
making it exact for a pair of off-shell
quarks. The matrix elements are instead computed
by projecting this phase space onto the ``nearest'' (in some sense) on-shell phase space. This
procedure relies upon the fact that the phase space is much more sensitive to the threshold
region than the matrix elements.

The third calculation, that we dubbed \bbfl{}, relies directly upon the full~\bbfourlshort{}
code in the \POWHEGRES{} framework~\cite{Jezo:2016ujg}.
In this generator also the matrix elements are generated
accounting for full off-shell effects, i.e. accounting for resonant and non-resonant contributions and preserving complete spin correlations.

The implementation of finite-width effects requires particular attention. A formula
for the full resummation of threshold-enhanced corrections accounting for
finite-width effects has been published a long time ago~\cite{Fadin:1987wz,Fadin:1990wx}. However, such formula is valid upon integration of the whole
final-state kinematics. At fixed invariant mass of the pair such kinematics is essentially frozen
if the top width is neglected. However, in order to develop a Monte Carlo generator
we need a formula that is valid also in the case when the top and
antitop decay off-shell. We carefully analyze this problem in
Appendix~\ref{app:NRfiniteGamma}, where we found the appropriate formula.
However, we did not attempt to implement it in the present work. This may be necessary
for a Monte Carlo description of $t{\bar t}$ production near threshold in context where
the threshold region is explored experimentally in much better detail, e.g. in an $e^+e^-$
context. We instead manipulated this formula with some approximations, in order to match it
with the result of ref.~\cite{Fadin:1987wz,Fadin:1990wx}, which is the formula that we actually
implemented. We believe that this is sufficient for the purposes of the present work, since,
as advocated in \nrr{} and confirmed by the current work,
finite-width effects are small in contexts where the resolution
of the top invariant mass is much larger than its width.

We used our calculations in order to try to clarify some of the issues discussed
in the introduction by focussing upon the cross section in the ATLAS bin, i.e. with
the requirement $\mtt<350$~GeV, and $p^*<50$~GeV, where $p^*$ is the top
momentum in the $t{\bar t}$ rest frame.
Our aim is to answer the following questions:
\begin{itemize}
\item Is the perturbative QCD calculation in the ATLAS bin, performed by adding
  threshold enhanced contributions up to the N$^3$LO order (as advocated in \nrr{})
  adequate to describe the corresponding
  cross section? Or are we already in a regime where the perturbative expansion diverges
  and should be resummed to all orders (using for example the GFRW method)?
\item Are finite-width effects really important in the ATLAS bin?
\item How large is the contribution of threshold enhanced effects over
  the baseline NLO results?
\item How large is the contribution that can be ascribed to ${t{\bar t}}$ bound states?
\end{itemize}

The paper is organized as follows. In Section~\ref{sec:thr1} we summarize the results of
\nrr{}, and document its extensions in the \thro{} generator. In Section \ref{sec:finitewidth}
we discuss the inclusion of finite-width effects in the \thrt{} generator. The implementation
of the phase space with off-shell top and its projection to the ``nearest'' on-shell
phase space is described there. We also discuss the corrections to the Lorentzian
describing the top and the $W$ decay due to the dependence of the width upon the
mass of the particle decay products. These corrections are included by default
in the \thrt{} generators, but can be switched off with an appropriate flag in
the {\tt powheg.input} file.

In Section~\ref{sec:bb4l} we describe the implementation of threshold corrections
in the \bbfl{} generator.
The analysis of what is really needed for a full
treatment of threshold enhanced effects when off-shell top production is only
briefly described in Section~\ref{sec:finitewidth0}, and a full discussion is
postponed to appendix~\ref{app:NRfiniteGamma}.

In Section~\ref{sec:numerics}
we examine the output of our generator. We begin by showing plots of the invariant mass
of the top pair $\mtt$ and discuss their features. In particular we examine an
interesting effect of cross section suppression below the nominal threshold when
the top virtuality is limited to a window below and above the nominal mass. This
effect reduces the bound state contribution when realistic cuts are used in the analyses.

In Section~\ref{sec:atlasbin} we present several tables of cross sections for the ATLAS
bin obtained with our generators at LO and NLO, also showing the result for a truncation
of the threshold expansion at fixed order. We separate out contributions that
can be ascribed to double top, single top and no top events.
We also show the effect of varying the scale of $\as$ used
for the evaluation of the threshold enhanced effects, and illustrate the
behaviour of the  perturbative expansion for the
threshold enhanced corrections in the colour singlet channel.

In Section~\ref{sec:garzelli} we compare our results with those of ref.~\cite{Garzelli:2024uhe},
obtained in Non-Relativistic-QCD at NLO. In that framework only fully inclusive cross
sections can be computed, and thus we compare their result with our full one, that
comprises double top, single top and no top events together.

In Section~\ref{sec:largeBinsSigma} we present tables of cross sections
obtained with different upper cuts on $\mtt$. As advocated in \nrr{}, the proper
scale for the evaluation of these cross section is the one associated with the
invariant mass cut, rather than a scale evaluated as a function of the invariant
mass on an event-by-event basis. Furthermore, we specify how large is the
cross section that can be ascribed to toponium production in the presence of these cuts.

In Section \ref{sec:Conc} we give our conclusions.

\section{Perturbative calculation neglecting finite-width effects}\label{sec:thr1}
In this section we discuss the perturbative calculation of the cross section along the
lines of NRR. For convenience we briefly summarise the basic results of that paper.
The spectral density of the $t{\bar t}$ system in the non-relativistic approximation is given by
\begin{eqnarray}
  \rho_l({\cal E}) &=& \theta(a_l) \times \frac{1}{\pi r_l^3} \sum_{n=1}^\infty \frac{1}{n^3} \delta({\cal E}-E_{l,n}) \label{eq:ttrho}
                +\frac{m^{3/2} }{4\pi^2}\sqrt{{\cal E}} F(b_lv^{-1}) \\
  F(z)&=& \frac{z}{1-\exp(-z)}, \label{eq:Fdef}
\end{eqnarray}
where ${\cal E}=\mtt-2m$, and $v=\sqrt{{\cal E}/m}$ is the velocity
of the quark in the $t{\bar t}$ rest frame.
The index $l$ is $1$ for the colour singlet and 8 for the colour octet state.
Furthermore
\begin{equation}
 r_l=\frac{2}{ma_l}, \quad E_{l,n}=-\frac{m}{4}\frac{a_l^2}{n^2},\quad b_l=\pi a_l,
\end{equation}
and 
\begin{equation}
  a_1=C_F\as, \quad a_8=-\frac{\as}{2N_C}.
\end{equation}
The factor $F(b_lv^{-1})$ is known as Sommerfeld factor~\cite{Sommerfeld,Sakharov}.
The main result of NRR is that eq.~(\ref{eq:ttrho}) has a perturbative expansion given by

\begin{equation}\label{eq:rhoFull}
  \rho_l({\cal E}) \to   \frac{1}{2\pi r_l^3}\sum_{n=1}^\infty \frac{1}{n^3}\delta({\cal E}-E_{l,n})+\frac{m^{3/2}}{4\pi^2}\sqrt{\cal E}
  F^+(b_lv^{-1}),
\end{equation}
where the $+$ notation on $F$ indicates that $F$ should be expanded in powers of its argument, which is proportional to $\as$, and
the expansion coefficients should be interpreted as distributions regulated according to analytic regularization.
Also the $\delta$ functions have arguments that depend upon $\as$ through the binding energy. However, they admit a Taylor expansion
in terms of their derivatives, and thus, consistently with the fact that the coefficients of the expansion can be interpreted as distributions,
the whole expression has a perturbative expansion in $\as$. The terms of the expansion that have been considered in NRR are summarised as
follows: one must replace the factor of $v$ in the Born cross section by
\begin{equation}
  v+\frac{b_l}{2}+\frac{b_l^2}{12v}+\frac{\zeta(3)}{4\pi^2}
  b_l^3  m\delta({\cal E}). \label{eq:enhancements}
\end{equation}
The phase space integral corresponds to an integration in ${\cal E}$. In the non relativistic
limit $\mathd {\cal E} \propto v\mathd v$, and thus we see that the first three terms
in the above expression are integrable, and do not need regularization.
In order to implement the above formula, starting from the \hvq{} implementation,
we compute the colour singlet and colour octet fraction of the Born cross section
in the Born term of the \hvq{} implementation, replacing the $v$ factor
in the Born term by
eq.~(\ref{eq:enhancements}) with $l=1$ or $8$ respectively.
Since the \hvq{} program does not allow for the integration of
distributions in the Born phase space, in order to handle the
$\delta$ function term, we replaced it by the following formula
\begin{equation}
  \delta({\cal E})\approx\frac{105}{16\Ecut^{7/2}}\sqrt{{\cal E}}(\Ecut-{\cal E})^2,\label{eq:approxdelta}
\end{equation}
whose integral in ${\cal E}$ is one.
The parameter $\Ecut$ is taken to be small, i.e. 1~GeV, and it is varied by a factor
of two above and below this value to check that the output does not depend sensibly upon it.
The above procedure is carried out at the level of the computation of the cross section,
with the $t$ and ${\bar t}$ mass fixed to the nominal top quark pole mass value. At this stage,
the \hvq{} phase space does not generate any event with $\mtt<2\mtop$, and this is why
we implement the contribution of the bound states above, rather than below, the nominal threshold.
This is a reasonable choice, since the squared matrix elements, once the phase space factor
$v$ has been removed, have a smooth behaviour near
threshold (see formulae~(2.3-2.6) in ref. NRR).

In ref. NRR, off-shell effects were introduced according to the original
method (briefly reviewed in sec.~\ref{sec:thr2}) employed in
the~\hvq{} generator, i.e. they were included while generating the
top decays in the partonic events before showering (i.e. when writing
the so-called ``Les Houches Event Files''~\cite{Alwall:2006yp}, LHEF
from now on).

\subsection{Spin singlet and triplet contributions}\label{sec:spinSinglet}
Besides the computation of the total cross section in the threshold bin, we also computed
its separation into its spin singlet and spin triplet component. This is achieved using the
variable $\cos\theta_{\ell^+\ell^-}$, that represents the angle between the lepton and antilepton directions in leptonic decays
of the $t{\bar t}$ system.
It is constructed by first boosting the top and antitop to the $t\bar{t}$
rest frame, and then computing the
angle of the lepton directions, which coincide with the top or antitop spin directions, with opposite signs
for the top relative to the antitop. The singlet component is distributed as $1 + \cos\theta_{\ell^+\ell^-}$
while the triplet behaves as $1 - \cos\theta_{\ell^+\ell^-}$. It is easy to see that
by multiplying the cross section by $(1+3\cos\theta_{\ell^+\ell^-})/2$ we extract
its spin singlet component.

\subsection {Higher order  threshold enhanced corrections.}
In order to gauge the convergence of the threshold expansion, we also computed some higher order terms.
Terms of order $(\as/v)^4$ arise from the expansion of the Sommerfeld factor
\begin{equation}
  F(z) = 1 + \frac{z}{2} + \frac{z^2}{12}-\frac{z^4}{720}+{\cal O}(z^6)
\end{equation}
where the first three terms are the ones considered so far,
and the first neglected term
is of order $z^4$, i.e. $(\as/v)^4$. In order to estimate its contribution,
we recall that it appears in expressions of the form
\begin{equation}\label{eq:intformv4}
  \int_0^{v_{\rm cut}} \mathd v  \, v^2 \left(\frac{1}{v^4}\right)_+ H(v^2),
\end{equation}
where $H$ is a smooth function of $v^2$. In the integrand of eq.~(\ref{eq:intformv4}) one factor
of $v$ arises from the $\sqrt{\cal E}$ term in front of the Sommerfeld factor, another power
of $v$ arises because $\mathd E \propto v \mathd v$. Expanding $H(v^2)$ in $v^2$ we get
\begin{equation}\label{eq:intformv5}
  \int_0^{v_{\rm cut}} \mathd v  \, v^2 \left(\frac{1}{v^4}\right)_+ (H(0)+{\cal O}(v^2)).
\end{equation}
The order $v^2$ reminder is integrable in $v$, irrespective of the $+$ prescription.
On the other hand, the term proportional to $H(0)$ is divergent, and we should use
analytic regularization to compute it. The basic feature of analytic regularization is
that all scaleless integrals of even powers of $v$ times the distribution over the whole
range $[0,\infty]$ should yield zero. In the case of interest we must have
\begin{equation}
  \int_0^\infty \mathd v  \, v^2 \left(\frac{1}{v^4}\right)_+ = 0.
\end{equation}
This could be achieved by replacing
\begin{equation}
  \int_0^{v_{\rm cut}} \mathd v  \, v^2 \left(\frac{1}{v^4}\right)_+ =
  -\int^\infty_{v_{\rm cut}} \mathd v  \, v^2 \left(\frac{1}{v^4}\right)_+.
\end{equation}
Rather than implementing this slightly cumbersome subtraction procedure in
our code, we have preferred to regularize the distribution, writing it as
\begin{equation}
  \left(\frac{1}{v^4}\right)_+ = \frac{1}{v^2}\left(\theta(v-\eta)\frac{1}{v^2}
    +\theta(\eta-v) P(v)\right),
\end{equation}
where $\eta$ is a tiny number and $P(v)$ is a polynomial such that
\begin{equation}
  P(\eta)-\frac{1}{\eta^2}=0,\quad
  \left[\frac{\mathd}{\mathd v} \left(P(v)-\frac{1}{v^2}\right)\right]_{v=\eta}=0
\end{equation}
and
\begin{equation}
  \int \mathd v \left(\theta(v-\eta)\frac{1}{v^2}+\theta(\eta-v) P(v)\right)=0\,.
\end{equation}
In practice we have used the form
\begin{equation}
  P(v)=-24\,\frac{v^3}{\eta^5}+45\,\frac{v^2}{\eta^4}-20\, \frac{v}{\eta^3}.
\end{equation}

Further terms of order $(\as/v)^5$ do arise from the expansion of the bound state delta functions in the
binding energy. The effect of these corrections will be discussed further in Section~\ref{sec:fullResum}.

It is not difficult to extend the reasoning of this section to the term of order $(\as/v)^6$ that arises
from the expansion of the Sommerfeld factor. We do not present the corresponding argument, but
we will estimate its numerical value in the following.


It is also instructive to examine the radius of convergence of the Sommerfeld
function. By inspecting eq.~(\ref{eq:Fdef}), we see that the closest pole is at
$z=2\pi i$, and thus the radius of convergence is $|z|=2\pi$. In our case
$z=C_F\as/v$, and $v$ in the ATLAS bin (i.e. for $\mtt<350\;$GeV)
is equal to 0.24. Since $\as \approx 0.14$
we are well inside the convergence circle.

\subsection{Estimate of the full resummation of the $(\as/v)^n$ series}\label{sec:fullResum}
It is also easy to estimate the effect of the inclusion of the threshold corrections to all orders.
We start from formula~(\ref{eq:ttrho}), and rewrite it in the following simplified form
\begin{equation}\label{eq:FULLSM}
  \rho_l({\cal E})=\frac{1}{4\pi^2}(m)^2\left[\theta(a_l) \times \frac{m b_l^3}{2\pi^2} \zeta_3 \delta({\cal E})
    +v F(b_lv^{-1}) \right]
\end{equation}
which differs by~(\ref{eq:ttrho}) only because of the replacement of each $\delta({\cal E}-E_n)$ with $\delta({\cal E})$
(we have also used the non-relativistic form for ${\cal E}=mv^2$). The effect of this approximation is easily checked
by shifting the argument of the delta function by an amount $E_1$, which is the largest binding energy. As anticipated
earlier, we have performed this check, and found that it yields negligible differences.
The numerical implementation of formula~(\ref{eq:FULLSM}) is achieved by replacing the factor of $v$ in the Born term
with the content of the square bracket in eq.~(\ref{eq:FULLSM}). The Sommerfeld factor is integrable, and the delta function
is treated with the same method used for eq.~(\ref{eq:enhancements}).

\section{Inclusion of finite-width effects}\label{sec:finitewidth}

\subsection{New implementation of top decays in POWHEG-hvq}\label{sec:thr2}
The \POWHEGhvq{} dates back to 2007, before the existence of the \POWHEGBOX{}
framework. The implementation of the top decay, based upon a concept published in~\cite{Frixione:2007zp},
also attempted to implement the production of off-shell top quarks. This was achieved by
transforming a given on-shell event into an off-shell one.
In doing so, an effort was made to include some known effects. For example, in the full off-shell propagator
the term proportional to $m\Gamma$ is really not constant, since it depends upon the virtuality.
The phase space for a pair of off-shell
top quarks, potentially with different virtualities, differs from the one for equal mass
on-shell tops, and a correction for this effect was attempted.
Similarly, by sending the top and antitop quark off their mass
shell, an on-shell to off-shell kinematic mapping is needed, and this also may alter the momentum
fraction of the incoming partons, so that some PDF reweighting was needed. These weights
where used for the generation of the off-shell configuration, but could not modify the
cross section for a given on-shell kinematics, since that was generated at an earlier stage
of the implementation. This approach worked well enough, but it is clear that it becomes
problematic for production near threshold, since in that region the mapping from on-shell to
off-shell kinematics becomes extremely delicate. We thus modified the generation of the off-shell configuration in \POWHEGhvq{}.

The new off-shellness implementation works as follows. The virtualities of the top quarks and
the $W$ bosons are generated just before the generation of the Born phase space, according of
the corresponding Lorentzian distribution, proportional to
\begin{equation} \label{eq:Lorentian}
  \frac{1}{\pi} \mathd S \frac{m\Gamma}{(S-m^2)^2+m^2 \Gamma^2}\,,
\end{equation}
where $m$ and $\Gamma$ stand for the pole mass and widths of the corresponding unstable particles. The generation
is limited by the following conditions:
\begin{eqnarray}
  m_b + h &<& S_{t/{\bar t}} < 2\mtop, \\
  h &<& S_{W^\pm}< 2m_W\,,
\end{eqnarray}
where $h$ is a technical parameter of the order of a typical hadronic scale, that we choose
equal to 2~GeV.
\subsection{Running-width corrections}\label{sec:runningGamma}
We optionally apply some corrections to the Lorentzian distributions, considering the fact that the decay
amplitude depends on the available phase space, and thus also on the virtuality of the
particle. We correct for this effects by multiplying the phase-space jacobian by a factor
\begin{equation}\label{eq:runningwidth}
  {  \frac{\sqrt{S}\tilde{\Gamma}(S)}{(S-m^2)^2+S \Tilde{\Gamma}^2(S)} }%
  \bigg/{  \frac{m\Gamma}{(S-m^2)^2+m^2 \Gamma^2} }
\end{equation}
for each resonance.
We recall that $m\Gamma$ is the squared amplitude for the decay (in fact
$\Gamma$ is equal to the squared amplitude divided by the mass). As such,
for an off-shell particle, it depends upon $S$.
In the case of the $W$, the decay amplitude must depend only upon
$S$, if we neglect the mass of the decay products. Thus on dimensional
ground, to account for running width effects, we must replace
\begin{equation}
  m\Gamma \Longrightarrow \frac{S}{m^2} \times m\Gamma,
\end{equation}
that with eq.~(\ref{eq:runningwidth}) implicitly defines
\begin{equation}
  \tilde{\Gamma}(S) = \frac{\sqrt{S}}{m}\Gamma.
\end{equation}
For the top, things are more delicate. The top width at leading order is given by
\begin{equation}
  \Gamma^{(0)}_t(\mtop,m_W,m_b)
  =\frac{G_{\rm F} \mtop^3 V_{tb}^2}{8\pi\sqrt{2}}\sqrt{(1-x_W-x_b)^2-4x_W x_b}
  \left((1-x_b^2)^2+x_W^2(1+x_b^2)-2x_W^4\right),
\end{equation}
where $x_W=m_W/\mtop$ and $x_b=m_b/\mtop$. One is tempted to substitute $\mtop$ and $m_W$ with their
virtualities to compute the correction, but this is incorrect.  This can be understood by
considering a very large top mass limit, such that the $W$ mass can be
neglected.
In this limit, the top decay is equivalent,
by the Goldstone-boson equivalence theorem,
to the decay into the charged Goldstone bosons
of the Higgs doublet, which in turn become the longitudinal
components of the $W^\pm$ after electroweak-symmetry breaking.

Two powers of the $\mtop^3$ appearing in
the width are then seen to arise from the Yukawa top coupling, and
only one power can arise from phase space.  Alternatively, when considering
the standard calculation of the width with the $W b$ final state, one can
see that two powers of $\mtop$ arise from violation of current
conservation in the terms where the $W$ momentum multiplies the axial top-bottom vertex,
that lead to a powers of $\mtop$ each. Thus, we define the correction factor for the top as
\begin{equation}
  {  \frac{\sqrt{S}\tilde{\Gamma}_t}{(S-m^2)^2+S \tilde{\Gamma}_t^2} }
  \bigg/{  \frac{m\Gamma_t}{(S-m^2)^2+m^2 \Gamma_t^2} },
\end{equation}
with
\begin{equation}
  \tilde\Gamma_t=\Gamma_t^{(0)}(V_t,V_W,m_b) \times \frac{\mtop^2}{V_t^2}
\end{equation}
where $V_t=\sqrt{S}$ and $V_W$ is the virtuality of the $W$.
The $m^2/V_t^2$ factor corrects two powers of the virtuality in $\Gamma^{(0)}_t$
that should have been instead powers of the top mass as discussed
above.
\subsection{On-shell mapping}

The phase space of the $t{\bar t}$ pair is generated with the top and
antitop mass set equal to the  $t$ and ${\bar t}$ virtualities generated as specified
above, and the Born phase space Jacobian is supplemented with the correction
factor that we just discussed.
By proceeding in this way, the phase space correctly accounts for the different
masses in the top and antitop, and the incoming momentum fraction already includes
off-shell effects.

Because of the structure of the \POWHEGBOX{} framework, the off-shell effects correctly propagate
to the full real phase space, which in the \POWHEGBOX{} is factorised into an underlying
Born phase space and a radiation phase space. However, the routines that compute
the Born, Real and Virtual contributions receive now as kinematics argument the
off-shell phase space, while the matrix elements are computed for on-shell particle.
At variance with the original \POWHEGhvq{} implementation, where we needed a mapping from the
on-shell kinematics to the off-shell one, now we need the opposite, i.e. a mapping from
the off-shell to the on-shell kinematics, so that we can compute the matrix elements.

A description of the on-shell to the off-shell mapping procedure was
never discussed in detailed in the original \POWHEGhvq{}
reference~\cite{Frixione:2007nw}. It was however reported in detail in
an appendix of the \ttMiNNLO{} publication~\cite{Mazzitelli:2021mmm}.
We report here the description of the inverse mapping scheme,
i.e. from the off-shell to the on-shell kinematics. The mapping has
the following properties:
\begin{itemize}
\item The mapping is performed on the phase space for the production of the
  $t{\bar t}$ pair, before their decay product is generated.
\item The mapping does not depend upon the $W$ virtualities; it depends only
  upon the $t$ and ${\bar t}$ ones.
\item The mapping does not affect final state particles other than the top
  and the antitop. Thus, in real graphs where radiation is present, the radiation
  is not affected by the mapping. Noticed that this feature also applies to ref.~\cite{Mazzitelli:2020jio}
  where the radiation of up to two light partons were considered.
\item In the rest frame of the $t{\bar t}$ pair the modulus of the spatial momentum of the
  $t$ and ${\bar t}$ are preserved by the mapping.
\end{itemize}
The mapping is then achieved through the following steps:
\begin{itemize}
\item We start with the partonic CM frame, which we denote with $F_{\rm PCM }$.
   We call $p_{t/{\bar t}}$ the top and antitop momenta in this frame.
   We boost the top and antitop  momenta with velocity
   \begin{equation}
     \vec{\beta} = - \frac{\vec{p}_t+\vec{p}_{\bar t}}{p_t^0+p_{\bar t}^0}.
   \end{equation}
   We call $p'_{t/{\bar t}}$ the top and antitop momenta after the boost. We have
   $\vec{p}'_t+\vec{p}'_{\bar t}=0$.
 \item We now define new top and antitop momenta $P'_{t/{\bar t}}$ as
   \begin{eqnarray}
     P'^0_{t/{\bar t}} &=& \sqrt{|\vec{p}'_{t/{\bar t}}|^2+\mtop^2}, \label{eq:newen}\\
     \vec{P}'_{t/{\bar t}}&=& \vec{p}'_{t/{\bar t}}.
   \end{eqnarray}
 \item We boost the $P'_{t/{\bar t}}$ momenta to the new top momenta  $P_{t/{\bar t}}$, using the velocity
   \begin{equation}
     \vec{\beta}' = \frac{\vec{p}_t+\vec{p}_{\bar t}}{\sqrt{(P'_t+P'_{\bar t})^2+(\vec{p}_t+\vec{p}_{\bar t})^2}}.
   \end{equation}
   Now we have
   \begin{equation}
     \vec{p}_t+\vec{p}_{\bar t}=\vec{P}_t+\vec{P}_{\bar t},
   \end{equation}
   so that with the new $P_{t/{\bar t}}$ the vector sum of the final state particle momenta is
   still zero in the $F_{\rm PCM}$ frame, but we now have $P_{t/{\bar t}}^2=\mtop^2$.
 \item The total energy of the final state is now no longer equal to the initial state one. We recover
   full momentum conservation by rescaling the initial parton momenta as
  \begin{equation}
    P_{1/2} =  \frac{E}{e} p_{1/2}
  \end{equation}
  where $e$ ($E$) are the total final state energy before (after) the transformation.
\end{itemize}
The mapping scheme illustrated above presents some technical problem in the implementation of the soft and
collinear matrix element in \POWHEG{}. We have thus preferred to adopt the following variant,
where eq.~(\ref{eq:newen}) is replaced by
\begin{equation}
  P'^0_{t/{\bar t}} = \frac{p'^0_t+p'^0_{\bar t}}{2}.
\end{equation}
With this new scheme, the final rescaling of
the initial momenta is unnecessary, but the mass used in the
calculation of the matrix elements is some sort of average of the top
and antitop virtualities, and thus differs in different events.

The on-shell mapping scheme that we have adopted required some
modification of the \POWHEGBOX{} file {\tt sigcollsoft2.f}. In
\POWHEG{} collinear and soft subtractions are computed in terms of the
Altarelli-Parisi splitting kernels and of the eikonal formula for soft
emission, convoluted or multiplied by the Born cross section. The
projection of real-emission configurations onto the underlying Born
kinematics in \POWHEG{} preserves the mass of the $t{\bar t}$ system,
and so does our on-shell projection scheme. Collinear singularities
only involve initial-state radiation, and are thus insensitive to the
top and antitop kinematics, since our projection scheme preserves
the four-momentum of the $t{\bar t}$ system. This is not the
case for the computation of soft subtraction terms, since the top and
antitop can also emit soft gluons. For instance, in the real matrix
elements terms of the form $P_t^2/(P_t\cdot k)$ (where $k$ is the
momentum of the radiated gluon) do arise in the soft limit. On the
other hand, the corresponding term in the soft counterterm of the
native \POWHEG{} implementation is $p_t^2/(p_t\cdot k)$,
thus spoiling soft cancellations.  Therefore we
modified the {\tt sigcollsoft2.f} file, making sure that the soft
limit is computed using the on-shell-projected kinematics ($P_{t/{\bar t}}$).

\subsection{The \bbfl{} generator}\label{sec:bb4l}
This generator is based upon the \bbfourlshort{} code~\cite{Jezo:2016ujg}.
Threshold corrections are computed by multiplying the Born cross
section by the threshold enhancement factor, obtained by assuming a 2/7
colour singlet and a 5/7 colour octet fraction.
In the NLO case, we correct the ${\cal O}(\as)$ term of the threshold
enhancement component by replacing its $\as$ factor with $\as - \as^{\rm (NLO)}$
where $\as$ stands here for the value of $\as$ used in threshold correction,
and $\as^{\rm (NLO)}$ stands for the value of $\as$ that the generator is using for the
NLO corrections. With this procedure we remove the over counting.

We make no attempt to separate the single and double resonant
contributions in the Born cross section. Such separation cannot be
performed in a gauge invariant way, and would be to a large extent
arbitrary.  We thus expect some unphysical enhancement of the single
top cross section near threshold, but, as we will see later, such
effects are only a small fraction of the single top cross section
contamination near the double resonance nominal threshold.  More
refined approaches are postponed to future work.

\subsection{Including finite-width effects}\label{sec:finitewidth0}
The inclusion of finite-width effects in a Monte Carlo framework is a delicate issue.
The full inclusion of threshold effects taking the finite width of the top into account
was carried out in refs.~\cite{Fadin:1987wz,Fadin:1990wx}. However, that calculation
is, strictly speaking, only valid when one integrates over the full phase space for the final state at fixed $t{\bar t}$
invariant mass. When the width is neglected, in proximity of the threshold only the $s$-wave
contribution counts, and the magnitude of the momenta of the top and antitop are fully
determined. This is not the case if off-shell effects are considered. In appendix~\ref{app:NRfiniteGamma}
we show how to modify the formalism in order to implement consistently finite-width effects
in a Monte-Carlo framework, i.e. also describing correctly the distributions of the top and antitop
virtualities. Such implementation
is mandatory for top pair production in a lepton-antilepton collider in order to accurately
describe the final state in the threshold region, and was actually carried out in ref.~\cite{Bach:2017ggt}.
As we argued in ref.~\cite{Nason:2025hix}, such precision is not actually needed at hadron colliders, since the threshold region is already smeared by detector effects.
A recipe is proposed in appendix~\ref{app:NRfiniteGamma} that is such that the prediction for the invariant
mass distribution of the $t\bar{t}$ system is correct when the energies and momenta of the
top and antitop are integrated over. We postpone to future work the
full implementation of the scheme detailed in appendix~\ref{app:NRfiniteGamma}.

\section{Numerical results}\label{sec:numerics}

In the following we will present results obtained with the following generators:
\begin{itemize}
\item[\thro{}]: This generator uses the \hvq{} program modified with the inclusion
  of threshold corrections computed in the limit $\Gamma_t\to 0$. The \POWHEGhvq{} decay mechanism
  generates the final state virtualities of the top and antitop, as detailed in section~\ref{sec:thr1}.
\item[\thrt{}]: We use the modified \hvq{} version described in
  section~\ref{sec:thr2}, with the correct phase space for the top and
  antitop virtualities. Threshold enhanced corrections are implemented
  including finite-width effects according to the procedure described in appendix~\ref{app:NRfiniteGamma}.
  In this generator we
  also allow for the inclusion of running width effects, as described
  in section~\ref{sec:runningGamma}. We will label as {\tt wc1} ({\tt wc0}) results
  where this option is on (off)
\item[\bbfl{}]: Here we start with the \bbfourl{} powheg implementation. Threshold corrections
  are again included using the recipe of appendix~\ref{app:NRfiniteGamma}.
\end{itemize}

We have generated event samples (at the LHEF level)
for proton-proton collisions at a centre of mass energy of 13~TeV,
and for a top pole mass of 172.5~GeV and a top width of 1.31 GeV,
setting the $W$ mass to 80.4 GeV with a width of 2.141 GeV.
The events include the leptonic
top decays into a pair of opposite sign electron. In the cross sections that we present, however,
we divide out the branching fraction, so that they can be interpreted as $t{\bar t}$ production
cross sections.
We use the {\tt NNPDF30\_nlo\_as\_0118} PDF set version 2 (LHAPDF ID = 260000)~\cite{NNPDF:2014otw}.
As usual, any set
included in the LHAPDF interface~\cite{Buckley:2014ana}, like the popular sets of
refs~\cite{Hou:2019efy,Bailey:2020ooq},
can be used transparently within the \POWHEGBOX{}, but we do not intend to perform a study on
PDF dependence here. All results are presented
for the default central scale choice for the hard process. This scale can be modified by acting upon the
{\tt facscfact} and {\tt renscfact} keywords in the {\tt powheg.input} file.

The scale choice for the strong coupling constant used in the
threshold correction factor needs to be set in a different
way. Ideally, it should be set equal to the absolute value of the
three-momentum flowing in the Coulomb propagator in the $t {\bar t}$
rest frame.  Since we do not have access to this variable (because we
are using the exact solutions of the Schr\"odinger equation for a
Coulomb potential) we proceed in a heuristic way as follows. The
binding energy of a bound $t {\bar t}$ system is equal to the square
of the Born radius divided by the top mass.  The inverse Born radius
is thus $\sqrt{m |E|}$, with $E=\mtt-2m$. Because of the finite width
of the top, we modify this formula as $\sqrt{m
  \sqrt{E^2+\Gamma_t^2}}$. Furthermore, we allow for a scale factor that
can be given as input to the program.  The scheme for the coupling of
the Coulomb interaction between heavy quarks is related to the
$\overline{\rm MS}$ scheme as
follows~\cite{Fischler:1977yf,Billoire:1979ih}
\begin{equation} \label{eq:alphaEff}
  \alpha_V(q)=\as(q)\left[1+\left(\frac{31}{3}-\frac{10}{9}n_f\right) \frac{\as(q)}{4\pi}\right].
\end{equation}
All the following analyses are performed at the level of Les Houches events, before the parton shower.
At this level, we identify the top quark as the $b\,l^+\,\nu_l$ system
and the antitop as the charge-conjugate system of decay products.
At the NLO level one extra parton is present in the event. The \thro{} and \thrt{} generators do not generate
radiation from the final state $b$ quark,
and thus we do not try to reconstruct $b$-jets by pairing the $b$ quark with the extra parton.
For the \bbfl{} generator at NLO radiation from the $b$ quark is present, and in this case
we reconstruct the $b$-quark jets using the anti-$k_t$ algorithm~\cite{Cacciari:2008gp} with a radius of $0.5$.

We separate the events in three categories:
\begin{itemize}
\item double top ({\tt dt}), where the mass of the top(antitop) decay products differs from the nominal top mass by less
  than 15~GeV.
\item single top ({\tt st}), where only one of the two masses differs from the nominal top mass by less
  than 15~GeV.
\item no top ({\tt nt}), where both the  top and antitop differ by more than 15 GeV from  the nominal top mass.
\end{itemize}

In the following we will refer to the results obtained at LO or NLO without the inclusion of threshold
correction as the ``baseline'' result. Furthermore, by LO or NLO we mean the accuracy of the generator
excluding the threshold corrections. Thus, LO/NLO baseline means the original generator at the LO/NLO level,
while LO/NLO full means the original generator at the LO/NLO level with the addition of threshold corrections
at the highest accuracy available.

\begin{figure}[htb]
  \begin{center}
    \includegraphics[width=0.8\textwidth,page=1]{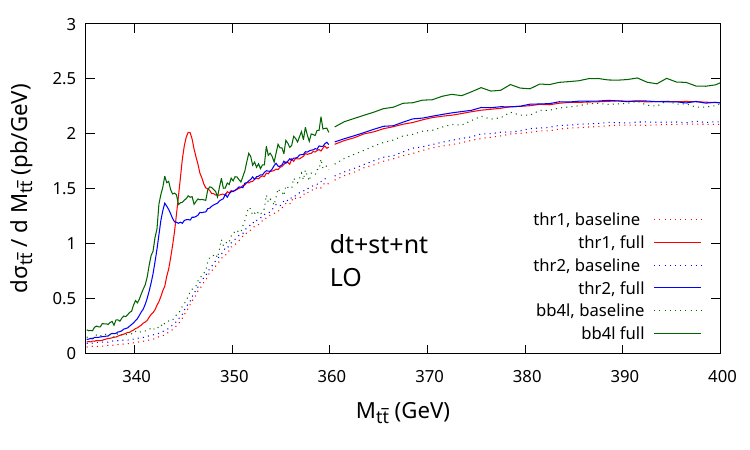}
  \end{center}
  \caption{\label{fig:mttfull} Invariant mass distribution
    of the $t\bar{t}$ pair. The double top, single top and no top contributions ({\tt dt}, {\tt st} and {\tt nt})
    are added together. The baseline cross section is obtained at the leading order level.
    The full cross section is obtained with the default \thro{}, \thrt{} and \bbfl{} setups.}
\end{figure}

We begin by showing in fig.~\ref{fig:mttfull}
the invariant mass distribution of the pair obtained with our three generators at the baseline LO
level. At the moment, the \thrt{} generator is used with the {\tt wc0} option (a comparison
to the {\tt wc1} option will be given later). Several observations are in order.

\begin{figure}[htb]
  \begin{center}
    \includegraphics[width=0.8\textwidth,page=8]{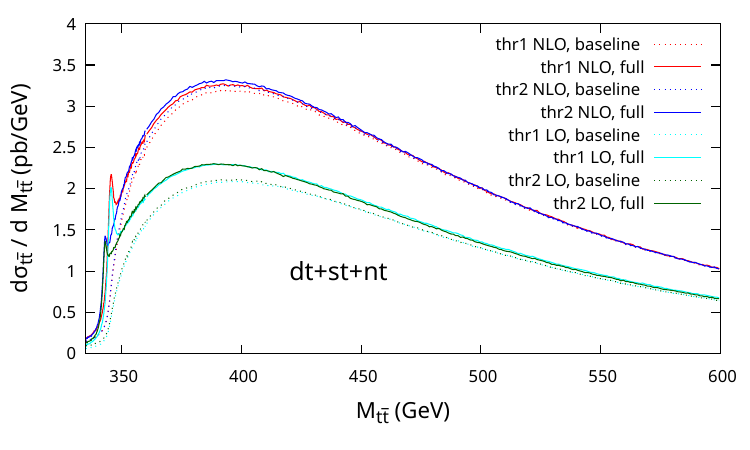}
  \end{center}
  \caption{\label{fig:mttfullNLO} As fig.~\ref{fig:mttfull}, for a larger
    range and including the NLO results, obtained with the default \thro{}, \thrt{}
    generators.}
\end{figure}

First of all,
we observe that the inclusion of Coulombic higher order corrections causes an increase of the cross
section even for very large $\mtt$ masses. This increase is easily found to arise from the
Coulomb contribution of order $\as$. At the NLO level such increase is
much less pronounced, since in that case the order $\as$ Coulomb
term must be subtracted to avoid overcounting. This is clearly demonstrated in fig.~\ref{fig:mttfullNLO},
for the \thro{} and \thrt{} generators,
where it is quite clear that the effect of threshold resummation is strongly reduced for large $\mtt$
at NLO. At the LO level the effect is larger, but there are no reasons to try to remove it, since
it is smaller than typical higher order corrections, and for non-relativistic top it represents
a true threshold correction. The \bbfl{} result, not shown in the figure, exhibits a similar behaviour

\begin{figure}[htb]
  \begin{center}
    \includegraphics[width=0.8\textwidth,page=2]{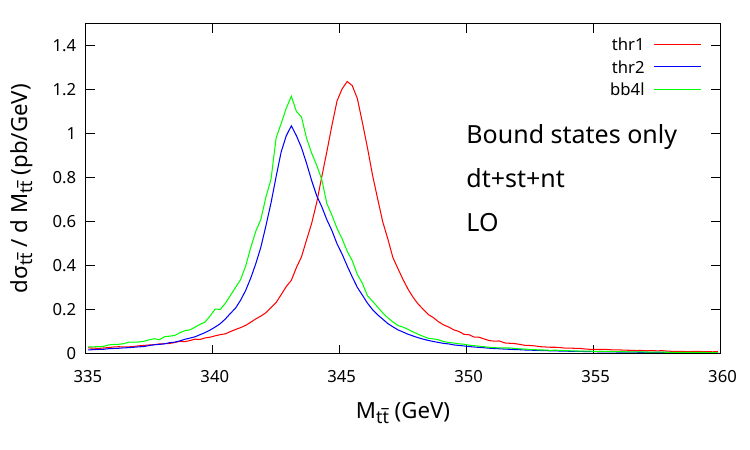}
  \end{center}
  \caption{\label{fig:mttBS} As in fig.~\ref{fig:mttfull}, showing only the bound
    state contribution to the invariant mass distribution
    of the $t\bar{t}$ pair. }
\end{figure}

A second observation regards the rather large effect of corrections near threshold.
In~fig.~\ref{fig:mttfullNLO} a peak associated with the (would be) toponium bound states is also visible, and in~fig.~\ref{fig:mttBS}
we show the contribution of the bound states alone, i.e. the cross
section due to eq.~(\ref{eq:approxdelta}) for the \thro{} generator, and to eq.~(\ref{eq:boundfinal}) for the \thrt{} and \bbfl{} generators.
The predictions of the \thrt{} and \bbfl{}
generators are in quite good agreement near threshold. As expected, the peak of the \thro{} generator is
displaced above the $2\mtop$ threshold, by construction.

\begin{figure}[htb]
  \begin{center}
    \includegraphics[width=0.8\textwidth,page=3]{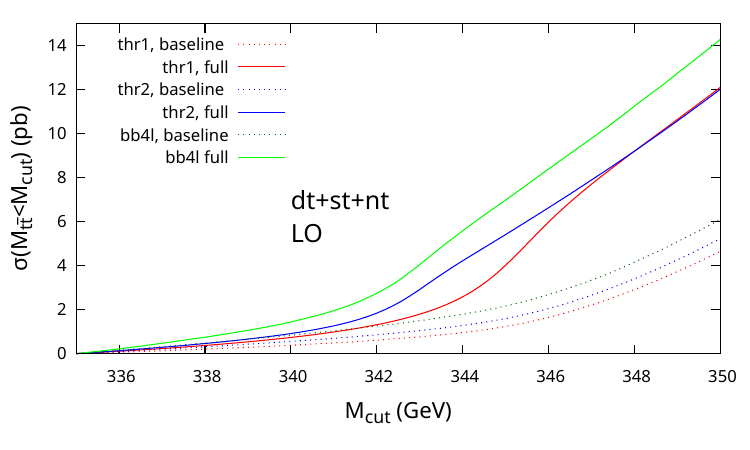}
  \end{center}
  \caption{\label{fig:mttcut} As in fig.~\ref{fig:mttfull}, showing the integrated
    cross section up to an invariant mass cut. }
\end{figure}

In fig.~\ref{fig:mttcut}
we show the integrated cross section as function of the invariant mass cut. In spite of the
differences found among the \thrt{} and \bbfl{} generators on one side and the \thro{} on the other,
the corresponding integrated cross section in the ATLAS region are  compatible in all three generators,
being in line with the differences in the baseline cross section.
This is consistent with the argument
of \nrr{}, where it was argued that the smearing due to finite-width effects was irrelevant for
cross sections measured with a coarse resolution.

\begin{figure}[htb]
  \begin{center}
    \includegraphics[width=0.8\textwidth,page=4]{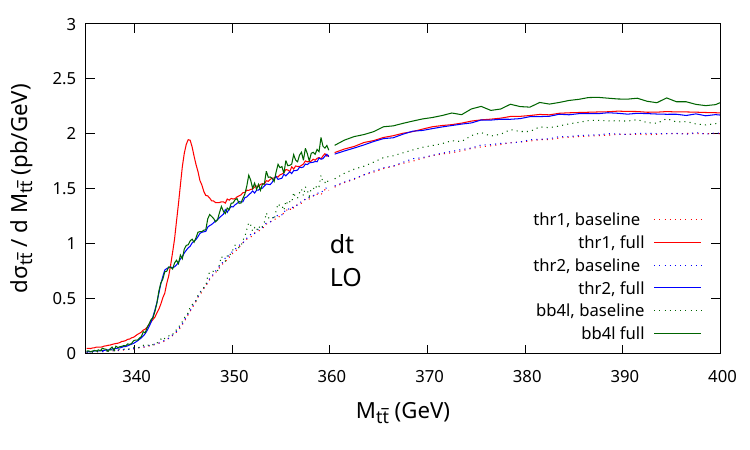}
  \end{center}
  \caption{\label{fig:mttfull-dt} As in fig.~\ref{fig:mttfull}, showing the contribution
    of the double top events. }
\end{figure}

\begin{figure}[htb]
  \begin{center}
    \includegraphics[width=0.8\textwidth,page=6]{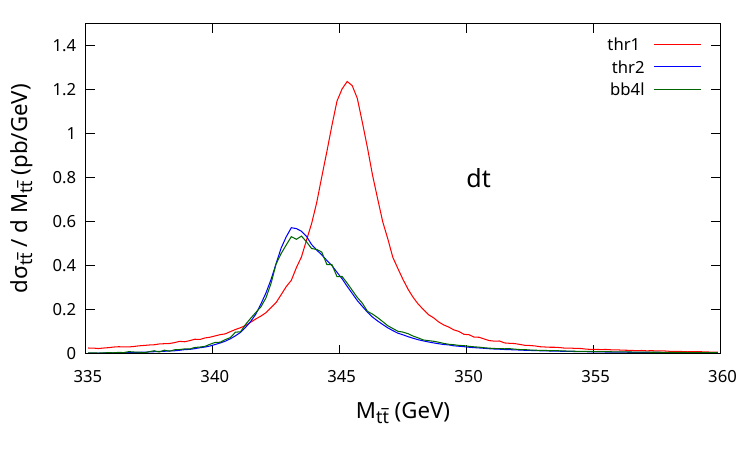}
  \end{center}
\caption{\label{fig:mttBS-dt} As in fig.~\ref{fig:mttBS}, showing only the
  contribution of the double top events.}
\end{figure}

\begin{figure}[htb]
  \begin{center}
    \includegraphics[width=0.8\textwidth,page=7]{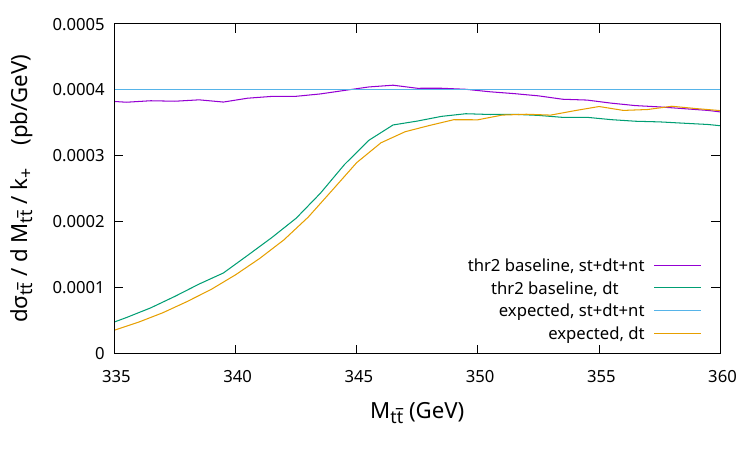}
  \end{center}
  \caption{\label{fig:phspaceplot} Invariant mass distribution divided by $k_+$ obtained with the
    baseline \thrt{} generator. The distributions labelled ``expected'' have been
    computed using formula~(\ref{eq:finalF0}) with a suitable normalization, both for the total
    inclusive cross section and for the double-top sample alone.}
\end{figure}

A different picture emerges if we consider the contribution of the double-top events alone,
as shown in fig.~\ref{fig:mttfull-dt}
and~\ref{fig:mttBS-dt}.
While the cross section of the \thro{} generator exhibits
the same behaviour of fig.~\ref{fig:mttfull}, the \thrt{} and \bbfl{} yield a smaller
bound states cross section.
We have investigated this behaviour at length, and have concluded that it arises
substantially from the phase space change associated with the double top mass constraint.
In fact, the dependence of the cross section upon the $\mtt$ mass in the threshold
region is essentially given by the phase space. From the result of eq.~(\ref{eq:finalF0}),
we see that the integrated phase space at fixed invariant mass of the pair
must be proportional to $k_+=\sqrt{(m/2) ( \sqrt{E^2+\Gamma_t^2}+E )}$, where $E=\mtt-2m$.
From fig.~\ref{fig:phspaceplot},
where we plot the invariant mass distribution divided by $k_+$, we can
see that for the full cross section this ratio is roughly constant,
while for the double-top restricted sample it decreases sharply below
the nominal threshold.  By implementing a numerical integration of the
phase space formula of eq.~(\ref{eq:finalF0}) we were also able to
verify the correctness of this behaviour from an independent
source. This suggests that, when restricting the top virtuality
to a window around the nominal mass, there is reduction in phase
space, that is bound to affect the cross section near the
resonances. As explained in appendix~\ref{app:NRfiniteGamma}, we have
not implemented the exact formula for the resummed cross section at
given top and antitop virtualities, and thus we cannot be absolutely
certain that this effects survives when a more exact treatment will become
available.  However, on physical ground, it is reasonable to assume
that when one of the two quarks is highly off shell it will not live
long enough to be affected by threshold effects, and thus the exact
formula for a highly off-shell quark should match the free formula,
and should be affected by the off-shell cuts in a similar way. In all
cases, although visible, the effect amounts to a reduction of the
cross section by an amount of the order of one picobarn.  We remark
that the \thro{} generator cannot implement this effect, since it uses
the full threshold enhanced corrections in the narrow width limit, and
off-shellness is generated later. This is one further reason to prefer
the \thrt{} (or the \bbfl{}) model over \thro{}.

\begin{figure}[htb]
  \begin{center}
    \includegraphics[width=0.8\textwidth,page=13]{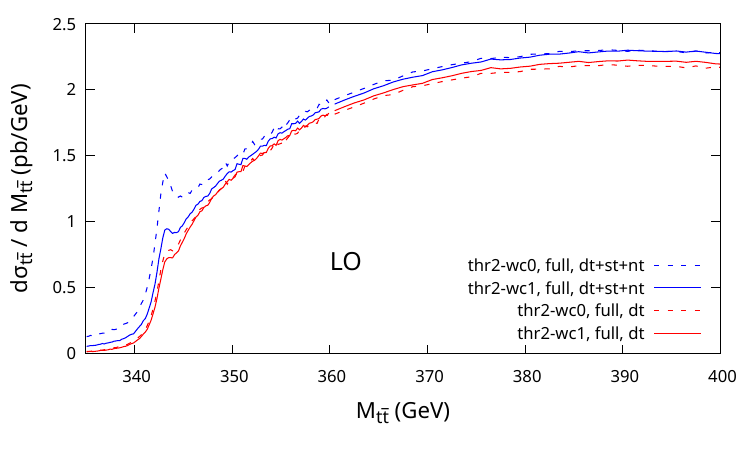}
  \end{center}
  \caption{\label{fig:wc0wc1} Comparison of the invariant mass distribution for the \thrt{}
    generator when running width effects are
    switched on ({\tt wc1}) or off ({\tt wc0}), for the fully inclusive cross section and for
    the double top sample alone.}
\end{figure}

When using the \thrt{} generator, we favour the inclusion of running width effects. In the plots
shown so far these effects were not included. In fig.~\ref{fig:wc0wc1}
we show this comparison. We see a rather dramatic effect for the fully inclusive sample,
while a very small effect is seen if only the double-top events are considered.
This effect is easily understood if one remembers that when running width effects
are included low virtualities for the top or the $W$ are suppressed, because
of the $\Gamma$ factor in the numerator of the Breit-Wigner distribution. This
strongly suppresses single top contributions, and, as a consequence, the full inclusive
sample, but not the double-top one.
\section{Tabulated results for the ATLAS bin}\label{sec:atlasbin}
In this section we present various cross sections for the ATLAS
bin. We show results for the \thro{}, \thrt{} and \bbfl{} generators,
showing also their expansion in perturbation theory up to the sixth
order. The contribution of the spin singlet channel, obtained following
the procedure detailed in sec.~\ref{sec:spinSinglet}, is also shown
separately. The spin triplet contribution can be obtained as the difference
from the total. We remind at this point that threshold corrections are spin
blind, and only depend upon the colour channel. The prevalence of the
spin singlet contribution near threshold is simply due to the well-known suppression
of the spin triplet component near threshold in gluon fusion~(see for example \nrr{}).

In table~\ref{tab:LOthr1thr2bb4l}
\begin{table}[htb]
   \begin{tabular}{|l|c|c|c|c|c|c|c|}
 \hline
 & LO & $+\as$ & $+\as^2$ & $+\as^3$ & $+\as^4$ & $+\as^6$ & Full \\
 \hline
 \multicolumn{8}{|c|}{\thro{} results} \\
 \hline
 \thro{} LO, dt & 3.764(5) & 4.978(6) & 7.93(1) & 10.25(1) & 11.40(1) & 11.1(1) & 11.18(1) \\ 
 \hline
 Spin 0 & 3.099(6) & 4.460(7) & 7.34(1) & 9.65(1) & 10.81(1) & 10.5(1) & 10.59(1) \\ 
 \hline
 \hline
 \thro{} LO, st & 0.553(2) & 0.664(2) & 0.838(3) & 0.930(3) & 0.968(3) & 0.98(3) & 0.961(3) \\ 
 \hline\hline
 \multicolumn{8}{|c|}{\thrt{} results without width corrections} \\
 \hline
 \thrt{} LO, dt & 3.852(6) & 4.876(6) & 7.103(9) & 8.57(1) & 9.10(1) & 8.48(1) & 8.32(1) \\ 
 \hline
 Spin 0 & 3.169(6) & 4.300(7) & 6.43(1) & 7.85(1) & 8.35(1) & 7.75(1) & 7.60(1) \\ 
 \hline
 \hline
 \thrt{} LO, st & 1.433(4) & 1.508(4) & 1.575(4) & 1.617(4) & 1.649(4) & 1.691(5) & 1.695(5) \\ 
 \hline
 \multicolumn{8}{|c|}{\thrt{} results including width corrections} \\
 \hline
 \thrt{} LO, dt & 3.769(5) & 4.777(6) & 6.953(9) & 8.37(1) & 8.84(1) & 8.19(1) & 8.06(1) \\ 
 \hline
 Spin 0 & 3.105(6) & 4.219(7) & 6.30(1) & 7.67(1) & 8.13(1) & 7.50(1) & 7.37(1) \\ 
 \hline
 \hline
 \thrt{} LO, st & 0.367(2) & 0.399(2) & 0.441(2) & 0.471(2) & 0.496(2) & 0.528(3) & 0.531(2) \\ 
 \hline
 \hline
 \multicolumn{8}{|c|}{bb4l results} \\
 \hline
 bb4l LO, dt & 3.99(2) & 5.05(3) & 7.28(4) & 8.71(4) & 9.20(5) & 8.66(4) & 8.52(4) \\ 
 \hline
 Spin 0 & 3.29(2) & 4.45(3) & 6.56(4) & 7.93(5) & 8.40(6) & 7.85(5) & 7.74(5) \\ 
 \hline
 \hline
 bb4l LO, st & 1.76(2) & 1.90(2) & 2.00(2) & 2.06(2) & 2.11(2) & 2.16(2) & 2.16(2) \\ 
 \hline
 \end{tabular}

  \caption{Cross sections in the threshold bin, obtained using the \thro{}, \thrt{} and \bbfl{} setups, including increasing orders of the
   Coulomb enhanced corrections, added to the basic leading order result.\label{tab:LOthr1thr2bb4l}}
 \end{table}
 we show the \thro{} and \thrt{} and \bbfl{} results at LO level in the ATLAS region.
We do not show in the table the {\tt nt} contributions, since they are all of the order of $10^{-2}\,$pb.
The \thrt{} results are shown with and without the widths corrections discussed in section~\ref{sec:thr2}.
First of all, we observe that the \thrt{} result with or without running width effects and the \bbfl{} ones
are in good agreement among each other. The table also show that the single top contribution
to the \thrt{} result with or without running width effects confirms the pattern observed and explained
in the previous section.

The \thro{} result yields cross sections that are larger by roughly
2.5~picobarns, and this effect is mostly due to the size of the $\as^3$
effects. As discussed in the previous section, we ascribe these larger corrections
to the fact that the \thro{} generator cannot account for the phase space
reduction arising when imposing the double-top constraint, and this is the
main reason why we favour the \thrt{} or the \bbfl{} result over the \thro{} one.

By looking at the perturbative expansion of the results, we see that including up to the $\as^3$ order already
captures most of the threshold enhanced effects, as already conjectured in \nrr{}. In all cases, the
spin 0 contribution dominates the cross sections.

We now discuss in more detail the convergence of the perturbative
expansion of the threshold enhanced contributions. This is better
done by restricting ourself to the colour singlet component of the
amplitude, which is the one that receives large, positive contributions
from enhanced threshold effects. In Table~\ref{tab:thrATLAS-LO-ColSing}
\begin{table}[htb]
  \begin{center}
 \begin{tabular}{|l|c|c|c|c|c|c|c|}
 \hline
 \multicolumn{8}{|c|}{\thro{}, colour singlet contribution} \\
 \hline
 & baseline & $+\as$ & $+\as^2$ & $+\as^3$ & $+\as^4$ & $+\as^6$ & Full \\
 \hline
 \thro{}, col. sing. & 0.833(1) & 2.998(4) & 5.805(8) & 8.13(1) & 9.28(1) & 9.2(1) & 9.053(9) \\ 
 \hline
 \end{tabular}

  \end{center}
  \caption{Colour singlet contribution to the cross section for $t{\bar t}$ production in the threshold bin, i.e. with
    $\mtt<350$~GeV, $p^\star<50$~GeV, at leading order, and including threshold enhanced
    corrections up to the order $\as^6$.}
  \label{tab:thrATLAS-LO-ColSing}
\end{table}
we show the cross section due to the colour singlet contribution only, restricting ourselves
for simplicity to the \thro{} case. We see from the table that baseline contribution is
less than a picobarn. This is not surprising, since in the dominant $gg$ channel and near threshold it amounts
to 2/7 of the total. The threshold enhanced contributions are very large for the first three orders, but
display a clearly convergent pattern when adding higher order terms, that contribute less than a picobarn
to the total. Thus, threshold enhanced corrections beyond the third order are quite
small, even for the ATLAS bin, that is very close to threshold.

In table~\ref{tab:NLOthr1thr2bb4l}
 \begin{table}[htb]
 \begin{tabular}{|l|c|c|c|c|c|c|c|}
 \hline
 & NLO & $+\as$ & $+\as^2$ & $+\as^3$ & $+\as^4$ & $+\as^6$ & Full \\
 \hline
 \multicolumn{8}{|c|}{\thro{} results} \\
 \hline
 \thro{} NLO, dt & 6.430(9) & 6.797(9) & 9.61(1) & 11.80(1) & 12.88(2) & 13.1(2) & 12.65(1) \\ 
 \hline
 Spin 0 & 5.07(1) & 5.43(1) & 7.88(2) & 9.82(2) & 10.78(2) & 11.0(2) & 10.58(2) \\ 
 \hline
 \hline
 \thro{} NLO, st & 1.424(4) & 1.476(5) & 1.756(5) & 1.907(6) & 1.969(6) & 1.97(4) & 1.957(6) \\ 
 \hline\hline
 \multicolumn{8}{|c|}{\thrt{} results without width corrections} \\
 \hline
 \thrt{} NLO, dt & 6.594(9) & 6.780(9) & 8.94(1) & 10.37(1) & 10.88(1) & 10.28(1) & 10.13(1) \\ 
 \hline
 Spin 0 & 5.25(1) & 5.42(1) & 7.27(1) & 8.50(2) & 8.93(2) & 8.40(2) & 8.27(1) \\ 
 \hline
 \hline
 \thrt{} NLO, st & 2.398(6) & 2.278(6) & 2.343(6) & 2.385(6) & 2.417(6) & 2.458(6) & 2.461(6) \\ 
 \hline
 \multicolumn{8}{|c|}{\thrt{} results including width corrections} \\
 \hline
 \thrt{} NLO, dt & 6.455(9) & 6.629(9) & 8.74(1) & 10.11(1) & 10.56(1) & 9.92(1) & 9.80(1) \\ 
 \hline
 Spin 0 & 5.16(1) & 5.32(1) & 7.13(1) & 8.31(2) & 8.70(2) & 8.13(2) & 8.03(1) \\ 
 \hline
 \hline
 \thrt{} NLO, st & 0.608(3) & 0.588(3) & 0.628(3) & 0.657(3) & 0.681(3) & 0.713(3) & 0.715(3) \\ 
 \hline
 \hline
 \multicolumn{8}{|c|}{bb4l results} \\
 \hline
 bb4l NLO, dt & 9.65(8) & 10.11(9) & 12.5(1) & 13.6(1) & 13.9(2) & 13.3(2) & 13.2(1) \\ 
 \hline
 Spin 0 & 7.9(1) & 8.4(1) & 10.7(1) & 11.8(2) & 12.1(2) & 11.5(2) & 11.4(2) \\ 
 \hline
 \hline
 bb4l NLO, st & 9.54(9) & 9.84(9) & 11.3(1) & 11.9(1) & 12.1(1) & 11.9(1) & 11.9(1) \\ 
 \hline
 \end{tabular}

   \caption{Cross sections in the threshold bin, obtained using the \thro{}, \thrt{} and \bbfl{} setups, including increasing orders of the
   Coulomb enhanced corrections, added to the basic NLO result.\label{tab:NLOthr1thr2bb4l}}
\end{table}
we display the same results at the NLO level.
We immediately notice that the $\as$ correction is very small. As anticipated earlier,
for such correction a subtraction was performed in the NLO case, thus leaving a small residual effect.
We also notice that the full prediction for the ATLAS bin is only marginally increased when going from the
LO to the NLO case. The pattern that we observe is similar to the LO case, with the noticeable exception
of the \bbfl{} result, that yields cross sections that are larger by roughly 3 picobarn with respect to the
\thrt{} ones. On the other hand, these three picobarns are already there at the baseline level. Furthermore,
we also see a large contribution of the single top sample, of the same size as the one of double tops.
We are thus inclined to interpret the larger \bbfl{} result as due to single top events leaking into
the double top sample. It is known that at the NLO level large corrections to the single top cross section
arise from the opening of new channels, as for instance due to bottom-gluon collision with the incoming
bottom arising from gluon splitting from one initial gluon. We also observe that, to some extent,
the relative magnitude of the {\tt st} and {\tt dt} \bbfl{} result at NLO might depend on the jet radius
we used to reconstruct the top and antitop quarks.

Finally, in table~\ref{tab:scaledep}
\begin{table}[htb]
   \begin{tabular}{|l|c|c|c|}
 \hline
 \multicolumn{4}{|c|}{Scale variations, baseline LO  } \\
 \hline
 Sc. fac.& 1 & 0.5 & 2 \\
 \hline
 \thro{} LO & 11.18(1) & 14.89(1) & 9.118(9) \\ 
 \hline
 \thrt{} LO  & 8.32(1) & 9.73(1) & 7.352(9) \\ 
 \hline
 \hline
 \bbfl{} LO  & 8.52(4) & 9.95(5) & 7.53(4) \\ 
 \hline
 \hline
 \end{tabular}

   \begin{tabular}{|l|c|c|c|}
 \hline
 \multicolumn{4}{|c|}{Scale variations, baseline NLO } \\
 \hline
 Sc. fac.& 1 & 0.5 & 2 \\
 \hline
 \thro{} NLO & 12.65(1) & 16.15(2) & 10.70(1) \\ 
 \hline
 \thrt{} NLO  & 10.13(1) & 11.49(1) & 9.18(1) \\ 
 \hline
 \hline
 \bbfl{} NLO  & 13.2(1) & 14.4(1) & 12.3(1) \\ 
 \hline
 \hline
 \end{tabular}

  \caption{Cross sections in the threshold bin for different
    choices of the scale factor used in the computation of the
    threshold corrections.\label{tab:scaledep}}
\end{table}
we show the effect of varying the scale used in the coupling constant
of threshold corrections by a factor of two below and above the central
value, which is detailed in the paragraph surrounding eq.~(\ref{eq:alphaEff}),
for the three generators. We notice that scale variation effects
are smaller for our favourite generators (i.e. \thrt{} and \bbfl{}), being
of the order of one picobarn.
This is presumably due to the fact that for these generators the $\as^3$
corrections are smaller because of the kinematic suppression due to
the cut in the top virtualities.

\section{Comparisons with available results}\label{sec:garzelli}
Ref.~\cite{Garzelli:2024uhe} presented a calculation of the total inclusive cross section
of $t{\bar t}$ production as a function of the invariant mass of the pair,
performed with the inclusion of threshold effects computed in the NLO-NRQCD framework,
and including the effect of soft gluon resummation. The NLO-NRQCD framework includes
all enhanced effects of order $(\as/v)^n$ and $\as (\as/v)^n$, and thus has higher precision
with respect to the calculation we have performed here.  Their result is totally
inclusive in the top and antitop virtualities, and thus cannot be used in a shower Monte Carlo context.
However, we are also able to compute fully inclusive cross sections,
and thus we can compare directly to their results, although some important limitations
have to be kept in mind:
\begin{itemize}
\item One of the Monte Carlo that they use is the \POWHEGhvq{} one. So, we can compare only to one
  of our generators that uses \POWHEGhvq{}.
\item We cannot use the \thro{} generator, since it
  neglects finite width effects.
\item Therefore we use the \thrt{} generator, although it uses an
  implementation of the phase space that allows us to go well below
  threshold, much more accurately than the standalone \POWHEGhvq{} one
  used in ref.~\cite{Garzelli:2024uhe}.
\item We assume that the lower cut of 340 GeV adopted in ref.~\cite{Garzelli:2024uhe} is
  likely to be necessary to avoid to push the \POWHEGhvq{} generator in a region where it
  is not reliable at all.
\end{itemize}
Thus, the comparison must be taken only as indicative.

The tables~\ref{tab:Garzelli-thr2-wc0} and \ref{tab:Garzelli-thr2-wc1}
show the comparisons.
\begin{table}[htb]
  {\small
    \begin{tabular}{|l||c||c||c|c||c|c||c|c||}
\hline
\multicolumn{9}{|c|}{\thrt{} NLO results, no widths corr.,   340~GeV$<m_{t{\bar t}}<\mcut$ } \\
\hline
 $\mcut$ & $\Delta$ from \cite{Garzelli:2026ctb} & Basel. & central &  $\Delta$ & low sc. & $\Delta$ & high sc. &  $\Delta$ \\
\hline 
345 & $ 2.73\pm 0.61 $ & 1.41(1) & 3.91(1) & 2.50(2) & 5.44(2) & 4.03(2) & 2.99(1) & 1.58(2) \\
\hline 
346 & $ 3.16\pm 0.75 $ & 1.95(1) & 5.24(1) & 3.29(2) & 6.91(2) & 4.96(2) & 4.20(1) & 2.25(2) \\
\hline 
347 & $ 3.48\pm 0.90 $ & 2.74(1) & 6.66(2) & 3.92(2) & 8.50(2) & 5.76(2) & 5.50(1) & 2.76(2) \\
\hline 
348 & $ 3.74\pm 1.08 $ & 3.78(1) & 8.21(2) & 4.43(2) & 10.20(2) & 6.42(2) & 6.94(1) & 3.16(2) \\
\hline 
349 & $ 3.97\pm 1.27 $ & 5.03(1) & 9.88(2) & 4.84(2) & 11.99(2) & 6.96(2) & 8.51(2) & 3.48(2) \\
\hline 
350 & $ 4.15\pm 1.47 $ & 6.48(1) & 11.67(2) & 5.20(2) & 13.90(2) & 7.43(2) & 10.23(2) & 3.75(2) \\
\hline 
351 & $ 4.31\pm 1.69 $ & 8.07(1) & 13.57(2) & 5.50(2) & 15.90(2) & 7.84(3) & 12.05(2) & 3.98(2) \\
\hline 
352 & $ 4.44\pm 1.92 $ & 9.79(2) & 15.58(2) & 5.78(2) & 18.00(2) & 8.21(3) & 13.99(2) & 4.19(2) \\
\hline 
353 & $ 4.56\pm 2.15 $ & 11.65(2) & 17.69(2) & 6.04(3) & 20.21(2) & 8.55(3) & 16.04(2) & 4.38(2) \\
\hline 
354 & $ 4.65\pm 2.4 $ & 13.61(2) & 19.89(2) & 6.27(3) & 22.48(2) & 8.87(3) & 18.17(2) & 4.56(3) \\
\hline 
355 & $ 4.72\pm 2.66 $ & 15.67(2) & 22.16(2) & 6.49(3) & 24.83(2) & 9.16(3) & 20.39(2) & 4.72(3) \\
\hline 
\end{tabular}

  }
  \caption{\label{tab:Garzelli-thr2-wc0} Cross section as a function of the upper $\mtt$
    cut, computed with the \thrt{} generator.
    The $\Delta$ represent the increase in cross sections
    when threshold corrections are added. The $\Delta$ results from
    ref~\cite{Garzelli:2026ctb} are reported in the first column
    for ease of comparison. The baseline column reports the cross section
    when no threshold corrections are added. The central, low scale and
    high scale results are obtained adding the full threshold corrections
    with the central scale, half the central scale and twice the
    central scale for the coupling used in the threshold corrections.
    Running width effects are not included.}
\end{table}
\begin{table}[htb]
  {\small
    \begin{tabular}{|l||c||c||c|c||c|c||c|c||}
\hline
\multicolumn{9}{|c|}{\thrt{} NLO results, widths corr.,   340~GeV$<m_{t{\bar t}}<\mcut$ } \\
\hline
 $\mcut$ & $\Delta$ from \cite{Garzelli:2026ctb} & Basel. & central &  $\Delta$ & low sc. & $\Delta$ & high sc. &  $\Delta$ \\
\hline 
345 & $ 2.73\pm 0.61 $ & 0.925(6) & 2.540(9) & 1.62(1) & 3.48(1) & 2.56(1) & 1.947(8) & 1.02(1) \\
\hline 
346 & $ 3.16\pm 0.75 $ & 1.366(7) & 3.58(1) & 2.22(1) & 4.64(1) & 3.27(1) & 2.895(9) & 1.53(1) \\
\hline 
347 & $ 3.48\pm 0.90 $ & 2.056(7) & 4.80(1) & 2.75(1) & 6.00(1) & 3.95(1) & 4.01(1) & 1.96(1) \\
\hline 
348 & $ 3.74\pm 1.08 $ & 3.005(8) & 6.19(1) & 3.18(1) & 7.52(1) & 4.51(2) & 5.30(1) & 2.30(1) \\
\hline 
349 & $ 3.97\pm 1.27 $ & 4.175(9) & 7.73(1) & 3.55(2) & 9.17(1) & 5.00(2) & 6.76(1) & 2.58(1) \\
\hline 
350 & $ 4.15\pm 1.47 $ & 5.53(1) & 9.39(1) & 3.87(2) & 10.94(2) & 5.42(2) & 8.35(1) & 2.82(2) \\
\hline 
351 & $ 4.31\pm 1.69 $ & 7.04(1) & 11.19(1) & 4.15(2) & 12.84(2) & 5.80(2) & 10.08(1) & 3.03(2) \\
\hline 
352 & $ 4.44\pm 1.92 $ & 8.69(1) & 13.10(2) & 4.41(2) & 14.84(2) & 6.14(2) & 11.92(1) & 3.22(2) \\
\hline 
353 & $ 4.56\pm 2.15 $ & 10.47(1) & 15.11(2) & 4.64(2) & 16.93(2) & 6.46(2) & 13.87(2) & 3.40(2) \\
\hline 
354 & $ 4.65\pm 2.4 $ & 12.36(1) & 17.22(2) & 4.86(2) & 19.12(2) & 6.76(2) & 15.92(2) & 3.56(2) \\
\hline 
355 & $ 4.72\pm 2.66 $ & 14.35(1) & 19.41(2) & 5.06(2) & 21.38(2) & 7.03(3) & 18.06(2) & 3.71(2) \\
\hline 
\end{tabular}

  }
  \caption{As in table~\ref{tab:Garzelli-thr2-wc0} for the \thrt{} generator
  with running width effects. \label{tab:Garzelli-thr2-wc1}}
\end{table}
The result from ref.~\cite{Garzelli:2024uhe} is the one in their table 1
that uses as baseline generator \hvq{}.
Overall, accounting for the errors, our results are fairly compatible with
those of ref.~\cite{Garzelli:2024uhe}

For our better generators, i.e. the \thrt{} including running width effects
the agreement is quite good also for the central
values. When running width effects are not included in our \thrt{} generator,
the comparison with ref.~\cite{Garzelli:2024uhe} favours instead
the use of a high scale.

In general, as the mass cut increases our result becomes larger than
that of ref.~\cite{Garzelli:2024uhe}. On the other hand, on the basis
of the considerations exposed in \nrr{}, we do expect that all three
Monte Carlo considered in this work would overestimate the cross
sections in large bins much above threshold. In fact, the analyticity
considerations exposed in \nrr{} indicated that the scale of the
coupling to be used in order to compute the inclusive cross section
for a given invariant mass cut should be the one associated with the
cut, rather than the one associated with each kinematic value of the
$\mtt$ mass that is being integrated over. This means that scale
compensating effects should arise at higher order in QCD, and these
effects should be present in the calculation of
ref.~\cite{Garzelli:2024uhe}. However, the bin used in
ref.~\cite{Garzelli:2024uhe} does not depend upon a single scale,
since the 340~GeV lower cut is also present.

\section{Cross section estimates for large bins}\label{sec:largeBinsSigma}
In order to give a more reliable estimate of the cross section in relatively large
bins defined by a single $\mtt$ upper bounds, in the spirit of ref~NRR
we implemented in our codes the possibility to run at a
fixed scale associated with a given invariant mass cut
$\mtt<\mcut$, equal to
$\sqrt{m\sqrt{(\mcut-2m)^2+\Gamma_t^2)}}$ and using then the same
prescription of formula~(\ref{eq:alphaEff}) for $\as$.
In table~\ref{tab:fixedScale}
\begin{table}[htb]
  \begin{center}
    \begin{tabular}{|l||c||c|c||c|c||}
\hline
\multicolumn{6}{|c|}{\thro{} NLO,  $\mtt< \mcut$} \\
\hline
 $\mcut$ & Baseline & Full & $\Delta$ & Full fixed scale &  $\Delta$ \\
\hline 
350 & 9.45(1) & 15.84(2) & 6.39(2) & 14.20(1) & 4.74(2) \\
\hline 
355 & 18.37(2) & 26.19(2) & 7.82(3) & 23.60(2) & 5.23(2) \\
\hline 
360 & 29.56(2) & 38.34(2) & 8.79(3) & 35.19(2) & 5.64(3) \\
\hline 
365 & 42.28(2) & 51.84(3) & 9.56(4) & 48.28(3) & 6.00(4) \\
\hline 
370 & 56.11(3) & 66.31(3) & 10.21(4) & 62.42(3) & 6.31(4) \\
\hline 
375 & 70.71(3) & 81.48(4) & 10.78(5) & 77.31(3) & 6.60(5) \\
\hline 
380 & 85.91(4) & 97.19(4) & 11.28(5) & 92.76(4) & 6.85(5) \\
\hline 
\end{tabular}

    \begin{tabular}{|l||c||c|c||c|c||}
\hline
\multicolumn{6}{|c|}{\thrt{} NLO, no widths corr.,  $\mtt< \mcut$} \\
\hline
 $\mcut$ & Baseline & Full & $\Delta$ & Full fixed scale &  $\Delta$ \\
\hline 
350 & 10.46(1) & 16.00(2) & 5.53(2) & 15.31(2) & 4.84(2) \\
\hline 
355 & 19.66(2) & 26.48(2) & 6.83(3) & 24.92(2) & 5.26(3) \\
\hline 
360 & 31.19(2) & 38.94(2) & 7.74(3) & 36.84(2) & 5.65(3) \\
\hline 
365 & 44.27(2) & 52.74(3) & 8.47(4) & 50.25(3) & 5.98(4) \\
\hline 
370 & 58.47(3) & 67.55(3) & 9.08(4) & 64.73(3) & 6.27(4) \\
\hline 
375 & 73.45(3) & 83.05(4) & 9.60(5) & 79.96(3) & 6.51(5) \\
\hline 
380 & 88.98(4) & 99.04(4) & 10.06(5) & 95.71(4) & 6.72(5) \\
\hline 
\end{tabular}

    \begin{tabular}{|l||c||c|c||c|c||}
\hline
\multicolumn{6}{|c|}{\thrt{} NLO, with widths corr.,  $\mtt< \mcut$} \\
\hline
 $\mcut$ & Baseline & Full & $\Delta$ & Full fixed scale &  $\Delta$ \\
\hline 
350 & 6.681(9) & 10.65(1) & 3.97(2) & 10.09(1) & 3.41(2) \\
\hline 
355 & 15.51(1) & 20.67(2) & 5.17(2) & 19.48(2) & 3.97(2) \\
\hline 
360 & 26.71(2) & 32.74(2) & 6.02(3) & 31.14(2) & 4.43(3) \\
\hline 
365 & 39.58(2) & 46.30(3) & 6.72(4) & 44.39(3) & 4.81(4) \\
\hline 
370 & 53.58(3) & 60.87(3) & 7.29(4) & 58.70(3) & 5.12(4) \\
\hline 
375 & 68.42(3) & 76.21(3) & 7.79(5) & 73.81(3) & 5.39(5) \\
\hline 
380 & 83.84(4) & 92.07(4) & 8.23(5) & 89.46(4) & 5.62(5) \\
\hline 
\end{tabular}

    \caption{Cross sections as a function of an invariant mass cut,
      with threshold corrections computed with a coupling associated
      with the scale of the invariant mass cut, for both the \thro{}
      and \thrt{} generator (including running width effects) at NLO.
      \label{tab:fixedScale}}
  \end{center}
\end{table}
we report a comparison of our standard results, obtained with the
scale for the threshold corrections given as a function of the invariant
mass of each event, with those obtained with such scale given as
a function of the mass upper cut. We see that already at $\mtt=350\,$GeV
the fixed scale result is smaller than the running scale one, by an amount
that is near 1.5~pb in the \thro{} case, and 0.5-0.7 in the \thrt{} case
depending upon whether running widths effects are included or not. These
differences raise to 4.3, 2.5 and 3.5 when the mass cut is at 380~GeV.
This reduction of the threshold effect is real, as discussed
at length in ref~NRR. Coulomb effects are perturbative, and thus
decrease when the scale raises. 
We remark, however, that for larger mass cuts these differences are a small
fraction of the overall cross section, and are smaller than the typical
perturbative uncertainties. Keeping this in mind, it makes sense to run the
generator using the default variable scales.

A further remark has to do with the relevance of finite-width effects
for the size of the threshold corrections. We see that the difference
between our zero-width approximation, i.e. the \thro{} generator,
with respect to \thrt{} with no running width effects,
which include mass corrections, is
below a picobarn for all mass cuts considered, as can be seen
from the last column of fig.~\ref{tab:fixedScale}. This was anticipated
in \nrr{}, since the finite width yield a smearing of the cross section,
and becomes irrelevant for bin sizes much larger than the width.
However, as we introduce effects that reduce the cross section
for top virtualities below the nominal threshold,
as we do when we include running width effects, we see differences
that go up to a little more than a picobarn. The same thing happens if we
introduce fiducial cuts on the top virtualities, as can be seen for
the ATLAS bin in table~\ref{tab:NLOthr1thr2bb4l}. Running width effects,
as well as fiducial cuts on the top virtualities,
have instead a smaller impact on the \thro{} result, as a consequence of
its poor modeling of off-shell effects near the nominal threshold.

\section{Bound state contributions to the cross section}
It is possible with our generator to compute the separate
contribution of the (would be) bound states to the cross sections.
In table~\ref{tab:bndsigma-fixedscale}
\begin{table}[htb]
  \begin{center}
    \begin{tabular}{|l||c||c|c||c|c||}
\hline
\multicolumn{6}{|c|}{\thro{} NLO,  $\mtt< \mcut$} \\
\hline
 $\mcut$ & Baseline & +Bnd & $\Delta$ & +Bnd fixed scale &  $\Delta$ \\
\hline 
350 & 9.45(1) & 13.96(1) & 4.51(2) & 12.43(1) & 2.97(2) \\
\hline 
355 & 18.37(2) & 23.15(2) & 4.79(2) & 20.99(2) & 2.62(2) \\
\hline 
360 & 29.56(2) & 34.42(2) & 4.87(3) & 31.95(2) & 2.39(3) \\
\hline 
365 & 42.28(2) & 47.19(3) & 4.91(4) & 44.52(2) & 2.24(3) \\
\hline 
370 & 56.11(3) & 61.05(3) & 4.94(4) & 58.24(3) & 2.13(4) \\
\hline 
375 & 70.71(3) & 75.67(3) & 4.96(5) & 72.75(3) & 2.05(5) \\
\hline 
380 & 85.91(4) & 90.89(4) & 4.98(5) & 87.89(4) & 1.98(5) \\
\hline 
\end{tabular}

    \begin{tabular}{|l||c||c|c||c|c||}
\hline
\multicolumn{6}{|c|}{\thrt{} NLO, no widths corr.,  $\mtt< \mcut$} \\
\hline
 $\mcut$ & Baseline & +Bnd & $\Delta$ & +Bnd fixed scale &  $\Delta$ \\
\hline 
350 & 10.46(1) & 14.21(1) & 3.75(2) & 13.71(1) & 3.24(2) \\
\hline 
355 & 19.66(2) & 23.52(2) & 3.86(2) & 22.44(2) & 2.78(2) \\
\hline 
360 & 31.19(2) & 35.10(2) & 3.90(3) & 33.73(2) & 2.53(3) \\
\hline 
365 & 44.27(2) & 48.20(3) & 3.93(4) & 46.64(3) & 2.37(4) \\
\hline 
370 & 58.47(3) & 62.41(3) & 3.94(4) & 60.72(3) & 2.25(4) \\
\hline 
375 & 73.45(3) & 77.40(3) & 3.95(5) & 75.61(3) & 2.16(5) \\
\hline 
380 & 88.98(4) & 92.95(4) & 3.96(5) & 91.07(4) & 2.09(5) \\
\hline 
\end{tabular}

    \begin{tabular}{|l||c||c|c||c|c||}
\hline
\multicolumn{6}{|c|}{\thrt{} NLO, with widths corr.,  $\mtt< \mcut$} \\
\hline
 $\mcut$ & Baseline & +Bnd & $\Delta$ & +Bnd fixed scale &  $\Delta$ \\
\hline 
350 & 6.681(9) & 9.30(1) & 2.62(1) & 8.92(1) & 2.24(1) \\
\hline 
355 & 15.51(1) & 18.24(2) & 2.73(2) & 17.48(2) & 1.97(2) \\
\hline 
360 & 26.71(2) & 29.48(2) & 2.77(3) & 28.53(2) & 1.81(3) \\
\hline 
365 & 39.58(2) & 42.37(2) & 2.79(3) & 41.29(2) & 1.71(3) \\
\hline 
370 & 53.58(3) & 56.38(3) & 2.80(4) & 55.20(3) & 1.63(4) \\
\hline 
375 & 68.42(3) & 71.23(3) & 2.81(5) & 69.99(3) & 1.57(4) \\
\hline 
380 & 83.84(4) & 86.65(4) & 2.82(5) & 85.35(4) & 1.52(5) \\
\hline 
\end{tabular}

  \end{center}
  \caption{\label{tab:bndsigma-fixedscale}
    Bound state contributions to the cross sections,
    also computed with the fixed scale associated with the upper bound
    of the $\mtt$ bin.}
\end{table}
we report such calculation for the \thro{} and \thrt{} generators with or without
running width effects. From the table we see that when we use the fixed scale
setup we get comparable contributions for the \thro{} and \thrt{} generator
with {\tt wc0}. The {\tt wc1} yields a smaller value, presumably due to the
suppression of the regions with far off-shell top quarks.
In the \thro{} case, the much larger contribution when the running scale is used
can be ascribed to the fact that as a consequence of our implementation of the
delta-type contribution we are using a scale of the coupling that is very close
to its minimum.

\section{Conclusions}\label{sec:Conc}
In \nrr{} some of us pointed out that toponium formation requires
times of the order of an inverse GeV, and if the measurement
resolution of the invariant mass of the top-antitop system is of the
order of tens of GeV we are not really sensitive to this phenomenon.
Cross section observables with such a coarse resolution can be computed
in perturbation theory at fixed order, and, as such, they are agnostic
regarding bound-state formation, which instead requires
an all orders resummation of the perturbative expansion.

One intuitive way to look at this problem is the following:
the production of the pair takes
place in very short time, and as such is calculable in perturbation
theory. The long time fate of the pair may require the resummation of
the perturbative expansion, but, by the uncertainty principle, we only
need this if we are able to measure the final state mass with enough
precision.

Whether we are sensitive or not to bound state formation, an
enhancement of the production cross section near threshold takes
place, and the QCD calculations of $t{\bar t}$ production at NLO and
NNLO include such effect. In \nrr{} the N$^3$LO threshold enhanced
correction was estimated in the narrow width limit, and it was argued
that even higher order corrections were negligible. It was also
conjectured that the finite width of the top quark has no significant
impact, as long as the invariant-mass resolution of the pair is much
larger than the top-quark width.

In the present work we have substantiated many of the claims of \nrr{} by
refining the calculation of threshold corrections.
We have added higher order contributions and explicitly taken into
account the effects of the finite top-width. To this end,
we have implemented
three event generators (accurate at the NLO level in QCD)
for $t{\bar t}$ production and decay that are capable to
include threshold-enhanced corrections to all orders in perturbation
theory. The first generator, that we dubbed \thro{}, uses the \POWHEGhvq{}
code as baseline, and adds threshold corrections neglecting finite
width effects. A second one, \thrt{}, uses a modified version of
\POWHEGhvq{}, where the phase space is computed accounting for
the top and antitop off-shellness. This captures the most important
finite-width effects near threshold. The third one, that we call \bbfl{}
uses the \bbfourl{} generator, and thus fully includes off-shell and
non-resonant contributions.

The \thrt{} and \bbfl{} generators implement the full resummation of
threshold enhanced contributions. This resummed result has been
known for a long time~\cite{Fadin:1987wz,Fadin:1990wx}. It yields
the cross section at fixed invariant mass of the pair near
threshold. In the zero-width limit the final state kinematics is fully fixed
by the invariant mass of the pair (keeping in mind that the $s$-wave
contribution is dominant, and thus there is no angular dependence upon the
top quark directions), but this is not the
case in the presence of off-shell effects.
A Monte Carlo generator, that should be
able to simulate the production of the pair of tops with off-shell
virtualities, needs a less inclusive result.

In Appendix~\ref{app:NRfiniteGamma} we have examined in detail the
requirements for the calculation of threshold enhanced effects at
fixed invariant mass of the pair and fixed masses of the top and antitop
decay products. In eq.~(\ref{eq:finalF}) we give
a formal expression that implements these requirements.

The actual implementation of formula~(\ref{eq:finalF}) is a demanding
task, therefore for the purpose of the present work, we have just used
our formal expression to better define an approximate recipe that can
be used in a Monte Carlo framework. In essence, the recipe is such
that the top and antitop can be produced with different virtualities
and that after integration over their virtuality one reproduces the
fully inclusive result of ref.~\cite{Fadin:1987wz}. As anticipated in
\nrr{} and demonstrated more explicitly in the present work, finite
width effects are small in the LHC context, and thus there is no need
to implement them in a very precise way. In contexts where the
invariant mass of the pair is accessed very precisely, like in the
$e^+e^-$ pair production of $t{\bar t}$
pairs~\cite{Strassler:1990nw,Hoang:1998xf,Hoang:2013uda,Beneke:2015kwa},
a Monte Carlo generator would need a full implementation of
formula~(\ref{eq:finalF}), that we leave for future work.\footnote{Our
  recipe is similar to one of the options recently proposed for the
  inclusion of threshold corrections in
  PYTHIA~\cite{Sjostrand:2026rcz}.  With respect to that work we have
  added a better understanding of what is missing in order to obtain a
  fully consistent result.}

The phenomenological studies that we have performed with our
generators confirm in full the conjectures of \nrr{}. We find that
even for the very narrow range of $\mtt<350\,$GeV, that is the region
that plays an important role in the ATLAS analysis
\cite{ATLAS:2026dbe}, the cross section enhancement is dominated by
the first three perturbative orders, as can be seen by examining
table~\ref{tab:thrATLAS-LO-ColSing}, where we show that the full cross
section differs by its truncation at order $\as^3$ by less than a
picobarn.

Regarding the effects of the finite width, we found differences that
can be above two picobarn. In this context, a few subtleties come into play
which are worth discussing. In table \ref{tab:fixedScale} we present
fully inclusive cross sections as functions of a cut on the maximum top
pair invariant mass.  When this cut equals 350~GeV, the total threshold
correction in \thro{} is very close to the one in \thrt{}-{\tt wc0}
(i.e. without running width effects). On the other hand, when
we add cuts on the virtualities of the top, the \thrt{} result is
considerably reduced below the nominal $t{\bar t}$ threshold.
This phenomenon is discussed at length in
section~\ref{sec:numerics} near figure~\ref{fig:mttfull-dt}.
We have observed this behaviour in our \thrt{} MC generators,
but we have also been able to confirm it in a semi-analytic way
(see the discussion near fig.~\ref{fig:phspaceplot}). We thus believe
that it is a real physical phenomenon, that cannot be reproduced
by the \thro{} generator. The dependence upon the implementation
of the running with corrections has a related explanation.
When running with effects are included, small virtualities of the
top quarks are suppressed. Thus, we see that the {\tt wc0} and {\tt wc1}
options lead to different results when the fully inclusive cross section
is computed, but only small differences if the fiducial cuts
on the mass of the top decay products are in effect.

We avoided quoting an explicit cross section of the
threshold enhanced effects in the present work. Single top production
does not suffer from threshold suppression, and thus becomes very important
and should be carefully accounted for, and this can only be done
in a full study of the signal and background simulation within
an experimental analysis framework. We instead provide
generators that can be used within such studies.
In particular, we recommend
the use of the default \thrt{} generator (that includes running width
effects) and the \bbfl{} one.
The \thro{} generator is only used here for the purpose of performing theoretical
studies and validation of the other two. We make it public so that anybody may
repeat our studies.
Our generators, by default, use a scale for the coupling entering threshold corrections
that is computed on an event-by-event basis, as a function of the invariant mass
of the pair.
We have also made clear, however, that
depending upon the observable one considers it may be preferable to
run our generators with a fixed scale, that is instead computed on the
basis of the invariant mass cut. Since the cross section increases as
the mass of the pair increases, the default Monte Carlo will tend to
overestimate the effect of threshold resummation in regions where they
are small fraction of the cross section.

In general, we can say that we expect
the cross section in the ATLAS bin to be near ten picobarn, and
the excess over the cross section computed with standard NLO
generators to be near four picobarns.
This numbers are compatible with what is quoted in
ref.~\cite{Garzelli:2024uhe}.
When using NNLO generator the excess should be further reduced by
an amount of the order of two picobarn, that is the typical size of the
threshold correction of relative order $\as^2$ (see
for example table~\ref{tab:LOthr1thr2bb4l}).

\section*{Acknowledgements}
We wish to thank Yu-Heng Yu, Otto Heinz Hindrichs and Regina Demina for
spotting a bug in the prerelease distribution of our code.
GP acknowledges support from the EU Horizon Europe research and innovation programme under the Marie-Sk{\l}odowska Curie Action ``POEBLITA - POlarised Electroweak Bosons at the LHC with Improved Theoretical Accuracy'' - grant agreement no. 101149251 (CUP H45E2300129000).
\clearpage
\newpage

\appendix

\section{Threshold limit in the finite width case.}\label{app:NRfiniteGamma}
\begin{figure}[htb]
  \begin{center}
    \includegraphics[width=0.5\textwidth]{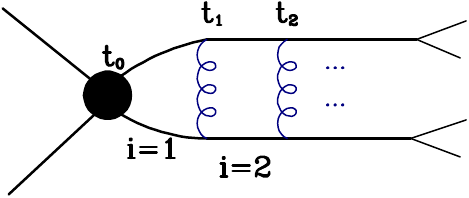}
    \end{center}
  \caption{\label{Fig:NRtt} Schematic representation of the Feynman graphs
    describing the production and decay of a $t{\bar t}$ pair
    near threshold.}
\end{figure}

Describing $t{\bar t}$ production in the threshold limit,
in the finite width case, requires some care. While the
approach of Ref.~\cite{Fadin:1987wz,Fadin:1990wx} is valid
in the inclusive case, when constructing a Monte
Carlo generator one should be able to take into account the fact
that the top and antitop may decay with different virtualities, and
the cross section should be computed for given virtualities of the
systems of decay products.
We thus examine $t{\bar t}$ production in the threshold region for given
virtualities of the top and anti-top. The relevant Feynman diagrams
are represented schematically in fig.~\ref{Fig:NRtt},
where the gluon exchange is to be interpreted as given only
by the Coulomb potential, which dominates the threshold
limit. We assume that the incoming vertex is pointlike, since
in the $t{\bar t}$ production amplitude the propagators are dominated
by momenta of the order of $\mtop$, much
larger than the energy involved in the threshold regime.

We must sum all graphs with 0 to infinity gluon rungs.
In the fermion propagators we must replace $m\to m-i\Gamma/2$, which arises from
the insertion of self-energy corrections.
In order to understand the structure of the formula, we transform the
Feynman graph to a non-relativistic perturbation theory graphs. This is done by
assigning a time $t_i$ to each vertex, and by replacing the associated
energy conservation delta function with its Fourier transform, i.e.
the integral over time of an
exponential of the sum of the energies entering the vertex multiplied
by $i$. The Coulomb
exchange is energy independent, and thus it is an equal time interaction.
We thus assign a single time to each Coulomb
interaction, denote as $t_i$ the time associated with the  $i^{\rm th}$
Coulomb exchange, and write the energy conservation delta function as
\begin{equation}
  2\pi \delta(e_i^{(t)}+e_i^{(a)}-e_{i+1}^{(t)} -e_{i+1}^{(a)})
  =\int \mathd t_i \exp(-i t_i(e_i^{(t)}+e_i^{(a)}-e_{i+1}^{(t)} -e_{i+1}^{(a)})),
\end{equation}
where by $e^{(t)}$ ($e^{(a)}$) we denote the kinetic energy
(i.e. the energy minus the mass)  of the top (antitop).
We will denote with
\begin{equation}
  e\equiv e^{(t)}+e^{(a)}
\end{equation}
their sum.

Our transformation to non-relativistic perturbation theory stops at the decay vertices, that we keep as they are in their Feynman graph formulation.

The energy of each propagator appears only in its adjacent vertices, and thus the energy
dependent factors can be collected as follows
\begin{equation}
  \frac{\exp(-i e_i^{(q)}(t_i-t_{i-1}))}{\left(k_i^{(q)}\right)^2-m^2+i\Gamma m} \approx
 \frac{1}{2m} \frac{\exp(-i e_i^{(q)}(t_i-t_{i-1}))}{e^{(q)}_i-\frac{\vec{k}_i^2}{2m}+i\frac{\Gamma}{2}},
\end{equation}
where $q$ stands for either $t$ or $a$.
On the right-hand side we have manipulated the denominator using the non-relativistic
approximation
\begin{equation}
  \left(k_i^{(q)}\right)^2-m^2+i\Gamma m = \left(e_i^{(q)}+m\right)^2 - \vec{ k}_i^2 -m^2+i\Gamma m
  \approx 2m e_i^{(q)} - \vec{ k}_i^2  +i\Gamma m.
\end{equation}
If $t_i>t_{i-1}$ the integral in $\mathd e_i^{(q)}/(2\pi)$ can be closed clockwise in the
lower complex plane, where the exponential damps the integrand,
leading to a contribution given by the pole at $e_i^{(q)}=\vec{k}_i^2/(2m)-i\Gamma/2$.
In the opposite case, i.e. $t_{i-1}>t_i$, the integral can be closed in the upper
complex plane, leading to zero. Putting the contributions of $q=t$ and $q=a$ together,
the result of the integration is
\begin{equation}\label{eq:int_e}
  \frac{1}{2m} (- i) \exp\left(-ie_i^{(q)}(t_i-t_{i-1})\right) \theta(t_i-t_{i-1}),
\end{equation}
where now
\begin{equation}
  e_i=\frac{\vec{k}^2}{m}-i\Gamma.
\end{equation}
Notice that the theta function will enforce strict time ordering, as is expected in
non-relativistic old-fashion perturbation theory.

The integrations in the times are done in sequence, starting with the earliest time, i.e. in our case $t_0$, the time of the production vertex. It enters
in the expression
\begin{equation}
  \exp(-i t_0(E-e_1)).
\end{equation}
Notice that the $-i\Gamma$ term in $e_1$ leads to a factor $\exp(t_0\Gamma)$,
so that the exponential is damped as $t_0\to -\infty$.
Integrating in $t_0$ from $-\infty$ to $t_1$ we get
\begin{equation}\label{eq:t0int}
 \frac{1}{-i} \frac{ \exp(-it_1(E-e_1))}{E-e_1},
\end{equation}
where the $-i$ in the denominator cancels the $-i$ in eq.~(\ref{eq:int_e}).

We now need to integrate in $t_1$ from $-\infty$ to $t_2$. The associated time
exponential is $\exp(-it_1(e_1-e_2))$, that combined with the exponential
in eq.~(\ref{eq:t0int}) yields $\exp(-it_1(E-e_2))$. The integral yields
\begin{equation}\label{eq:t1int}
\frac{ \exp(-it_2(E-e_2))}{E-e_2},
\end{equation}
It is easy to see that adding further rungs and integrating in time up
to the next time we always get energy denominators of the form $1/(E-e_i)$.
When we reach the last rung, we are left with the exponential factor
$\exp(-it_n(E-E_t-E_{\bar t}))$, where now the $t_n$ integral is unrestricted,
and $E_t$, $E_{\bar t}$ are the energies flowing in the outgoing top and
antitop propagators, and equal the energies of their respective decay
products. We thus recover the energy conservation delta function
$2\pi\delta(E-E_t-E_{\bar t})$.

At each intermediate state we must also integrate in the space momentum $\vec{k}$.
Since we must sum over the repetition of an arbitrary number of rungs, we use an operator notation,
and define operators in $\vec{k}$ space with the product defined as
\begin{equation}
  (A B)(\vec{k},\vec{k}'') =\int \frac{\mathd^3\vec{k'}}{(2\pi)^3} A(\vec{k}, \vec{k}')B(\vec{k}',\vec{k}'').
\end{equation}
We introduce the operators
\begin{eqnarray}
  T(\vec{k},\vec{k'})&=&{\mathbf 1}(\vec{k},\vec{k}') (E-e(k)), \\
  V(\vec{k},\vec{k'})&=&\frac{\alpha}{|\vec{k}-\vec{k'}|^2},
\end{eqnarray}
where
\begin{equation}
  {\mathbf 1}(\vec{k},\vec{k'})=(2\pi)^3 \delta^3(\vec{k}-\vec{k'}).
\end{equation}
The sum over all rungs can be then written in compact notation as
\begin{equation}
  S=\sum_{i=0}^\infty (T^{-1}V)^i.
\end{equation}
We must have
\begin{eqnarray}
  S&=&\mathbf{1}+T^{-1}V S, \nonumber \\
  (\mathbf{1}-T^{-1}V)S&=&\mathbf{1} \nonumber \\
  S&=&(\mathbf{1}-T^{-1}V)^{-1}=\left(T^{-1}(T-V)\right)^{-1}=(T-V)^{-1} T.
\end{eqnarray}
We now notice that
\begin{equation}
  (T-V)^{-1}=\frac{1}{E-H} \equiv R,
\end{equation}
where
\begin{equation}
  H(\vec{k},\vec{k'})\equiv {\mathbf 1}(\vec{k},\vec{k'}) e(\vec{k})+V(\vec{k},\vec{k'})
\end{equation}
is the non-relativistic Hamiltonian, and $R$ is the resolvent.
We can compute it in the basis of the eigenvectors of the Hamiltonian.
Our result can be written schematically as
\begin{equation}
  \int \frac{\mathd^3 \vec{k}_1}{(2\pi)^3} R(\vec{k}_1,\vec{k}_f) \frac{\left(E-\frac{\vec{k}_f^2}{m}
    +i\Gamma\right)2\pi\delta(E-E_{\bar t}-E_t)}{(2m)^2\left(E_t-\frac{\vec{k}_f^2}{2m}+i\frac{\Gamma}{2}\right)
    \left(E_{\bar t}-\frac{\vec{k}_f^2}{2m}+i\frac{\Gamma}{2}\right)}
  \times \mbox{decay amplitudes}.
\end{equation}
This is the amplitude, excluding the decay amplitudes of the top quarks. By squaring
and supplying the remaining decay amplitudes we get
\begin{eqnarray}
&& \int \frac{\mathd E_t}{2\pi}\frac{\mathd E_{\bar t}}{2\pi}\frac{\mathd^3 \vec{k}_f}{(2\pi)^3}  |R(\vec{x}=0,\vec{k}_f)|^2 \frac{\left(\frac{\Gamma}{2m}\right)^2\left(\left(E-\frac{\vec{k}_f^2}{m}\right)^2+\Gamma^2\right)2\pi \delta(E-E_t-E_{\bar t})}
  {\left(\left(E_t-\frac{\vec{k}_f^2}{2m}\right)^2+\frac{\Gamma^2}{4}\right)
    \left(\left(E_{\bar t}-\frac{\vec{k}_f^2}{2m}\right)^2+\frac{\Gamma^2}{4}\right) } \nonumber \\
  &\times& \frac{D((E_t+m)^2-\vec{k}_f^2)}{2m\Gamma}  \frac{D((E_{\bar t}+m)^2-\vec{k}_f^2)}{2m\Gamma}
           \label{eq:finalF}
\end{eqnarray}
where $\vec{k}_f$ is the momentum after the last Coulomb exchange, i.e. after the
$n^{\rm th}$ Coulomb exchange for each term of the series.
Eq.~(\ref{eq:finalF}) is the exact formula that one should use for the resummation of threshold
effects accounting for off-shell effects for the top and antitop in the final state.
It was actually implemented in ref.~\cite{Bach:2017ggt} in the framework of $e^+e^-$ collisions.

The $R$ dependent factor is a function of $E$ and $k_f=|\vec{k}_f|$, and is  given by
\begin{equation}
  R(\vec{x}=0,\vec{k}_f) = \sum_n \psi_n(0)\frac{1}{E-e_n+i\Gamma/2}\tilde\psi_n(|\vec{k}_f|)
  +\int \mathd E \psi_e(0)\frac{1}{E-e+i\Gamma/2}\tilde\psi_e(|\vec{k}_f|),
\end{equation}
where $\psi_{n/e}(x)$ and $\tilde\psi_{n/e}(k)$ are eigenstate of the Hamiltonian in the position and momentum
space representation respectively.

The $D$ factors are the squared decay amplitudes. They are nearly insensitive to $E$ and $\vec{k}_f$ near threshold, and after integration yield $2m\Gamma$. So, the energy integral yields
\begin{equation}
  \int \frac{\mathd E_t}{2\pi}\frac{\mathd E_{\bar t}}{2\pi}  \frac{
    \left(\frac{\Gamma}{2m}\right)^2\left(\left(E-\frac{\vec{k}_f^2}{m}\right)^2+\Gamma^2\right)2\pi \delta(E-E_t-E_{\bar t})}
  { \left(\left(E_t-\frac{\vec{k}_f^2}{2m}\right)^2+\frac{\Gamma^2}{4}\right)
    \left(\left(E_{\bar t}-\frac{\vec{k}_f^2}{2m}\right)^2+\frac{\Gamma^2}{4}\right) }
  =  \frac{\Gamma}{2 m^2}.
\end{equation}
We are thus left with
\begin{eqnarray}
\frac{\Gamma}{2m^2}\int \frac{\mathd^3 \vec{k}_f}{(2\pi)^3}  |R(\vec{0}_x,\vec{k}_f)|^2
  &=&\frac{\Gamma}{2m^2}\langle 0| \frac{1}{E-H}\frac{1}{E-H^*}| 0 \rangle  \nonumber \\
  &=& \frac{\Gamma}{2m^2}\sum_n \langle 0| \frac{1}{E-H}\frac{1}{E-H^*} |\psi_n\rangle\langle \psi_n| 0 \rangle \nonumber \\
  &=&\frac{\Gamma}{2m^2}\sum_n \langle 0|\psi_n\rangle \frac{1}{(E-E_n)^2+\Gamma^2}\langle \psi_n| 0 \rangle \nonumber \\
     &=& \frac{\Gamma}{2m^2}\sum_n | \psi_n(0)|^2 \frac{1}{(E-E_n)^2+\Gamma^2},
     \label{eq:finalF1}
\end{eqnarray}
which is the starting formula that leads to the result of ref.~\cite{Fadin:1987wz}.
We have not accounted fully for the normalization factors in our derivation, but
we can fix it now by contrasting formula~(\ref{eq:finalF})
with the analogous formula when no Coulomb exchanges are considered. It is given by
\begin{eqnarray}
  && \int \frac{\mathd E_t}{2\pi}\frac{\mathd E_{\bar t}}{2\pi}\frac{\mathd^3 \vec{k}_f}{(2\pi)^3} \frac{\left(\frac{\Gamma}{2m}\right)^2
     2\pi \delta(E-E_t-E_{\bar t})}%
  { \left(\left(E_t-\frac{\vec{k}_f^2}{2m}\right)^2+\frac{\Gamma^2}{4}\right)
    \left(\left(E_{\bar t}-\frac{\vec{k}_f^2}{2m}\right)^2+\frac{\Gamma^2}{4}\right) } \nonumber \\
  &\times& \frac{D((E_t+m)^2-\vec{k}_f^2)}{2m\Gamma}  \frac{D((E_{\bar t}+m)^2-\vec{k}_f^2)}{2m\Gamma}.  \nonumber \\
  &=&\frac{\Gamma}{2m^2}\left[\int \frac{\mathd^3 \vec{k}_f}{(2\pi)^3}
\frac{1}{\left(E-\frac{\vec{k}_f^2}{m}\right)^2+\Gamma^2
}\right] = \frac{1}{8 \pi\sqrt{2m}} \sqrt{\sqrt{E^2+\Gamma^2}+E} .  \label{eq:finalF0}
\end{eqnarray}
Thus, we
introduce the following recipe. We multiply our cross section by the factor
\begin{equation}
  \frac{\sum_n | \psi_n(0)|^2 \frac{1}{(E-E_n)^2+\Gamma^2}}
  {\left[\sum_n | \psi_n(0)|^2 \frac{1}{(E-E_n)^2+\Gamma^2}\right]_{\as=0}}.
\end{equation}

The starting formula for the spectral density is
\begin{equation} \label{eq:rhodef}
  \rho =  \sum_i |\psi_i(0)|^2 \delta(E_i-E)
\end{equation}
where $\psi_i(x)$ and $E_i$ are the eigenvectors and eigenvalues of the Hamiltonian. Formula (\ref{eq:rhodef}) gives rise to formula~(\ref{eq:ttrho}) in the toponium case. In order to simplify the notation we omit to indicate with a suffix $l$ the singlet and octet contribution. The formula
that we obtain works for any sign of the coupling constant.

If we want to include the effects of the finite top lifetime, we must replace
\begin{equation}
  \delta(E_i-E) \to \frac{1}{\pi} \frac{\Gamma}{(E-E_i)^2+\Gamma^2},
\end{equation}
where $\Gamma$ is the width of the top The width of the $t{\bar t}$ system is thus $2\Gamma$, that explains why
the $\gamma^2$ term is not divided by four in the above equation. This leads to the result
\begin{eqnarray}
  \rho&=& \rho_{\rm cont.}+ \rho_{\rm bound},  
          \\
  \rho_{\rm bound}&=&\theta(r_l)\sum_n \frac{1}{\pi^2 n^3 r_l^3} \frac{\Gamma}{(E-E_n)^2+\Gamma^2}   \label{eq:boundfinal}
  \\
  \rho_{\rm cont.}&=&\frac{4}{(2\pi)^3} \int k^2 \mathd k
  \frac{\frac{2\pi}{kr_l}}{1-\exp\left(-\frac{2\pi}{kr_l}\right)}
  \frac{\Gamma}{(E-E_k)^2+\Gamma^2}.\label{eq:cont}
\end{eqnarray}
After some algebraic manipulations we get
\begin{equation} \label{eq:contfinal}
  \rho_{\rm cont.}= \frac{m^2}{4\pi^2}\left[ \frac{k_+}{m}+\frac{2k_1}{m} \arctan\left(\frac{k_+}{k_-}\right)
    +\sum_{n=1}^\infty \frac{2k_1^2}{n^4 m^2}\frac{-|k_1|\Gamma n+k_+\left(n^2\sqrt{E^2+\Gamma^2}+k_1^2/m\right)}{\left(E+\frac{k_1^2}{m n^2}\right)^2+\Gamma^2}\right]
\end{equation}
where
\begin{equation}
  k_1=\frac{1}{r},\quad\quad k_\pm=\sqrt{\frac{m}{2}(\sqrt{E^2+\Gamma^2}\pm E)}.
\end{equation}
Combining eq.~(\ref{eq:contfinal}) with eq.~(\ref{eq:boundfinal}), and using the relation $2 k_1\theta(k_1)-|k_1|=k_1$  we get
\begin{equation} \label{eq:rhofinal}
  \rho= \frac{m^2}{4\pi^2}\left[ \frac{k_+}{m}+\frac{2k_1}{m} \arctan\left(\frac{k_+}{k_-}\right)
    +\sum_{n=1}^\infty \frac{2k_1^2}{n^4 m^2}\frac{k_1\Gamma n+k_+\left(n^2\sqrt{E^2+\Gamma^2}+k_1^2/m\right)}{\left(E+\frac{k_1^2}{m n^2}\right)^2+\Gamma^2}\right],
\end{equation}
which is the formula of refs.~\cite{Fadin:1987wz,Fadin:1990wx}. For numerical evaluation, and also to better display
the first few perturbative orders, it is convenient to rewrite it in the following form
\begin{eqnarray} \label{eq:rhofinalform}
  \rho&=& \frac{m^2}{4\pi^2}\Bigg\{ \frac{k_+}{m}+\frac{2k_1}{m} \arctan\left(\frac{k_+}{k_-}\right) \nonumber \\
      &+& \frac{2k_1^2}{m^2}\frac{k_1\Gamma \zeta(3)+k_+\left(\zeta(2)\sqrt{E^2+\Gamma^2}+\zeta(4) k_1^2/m\right)}{E^2+\Gamma^2} \\
      &+&\sum_{n=1}^\infty \frac{2k_1^2}{n^4 m^2}\left(k_1\Gamma n+k_+\left(n^2\sqrt{E^2+\Gamma^2}+k_1^2/m\right)\right)
          \times \Bigg[\frac{1}{\left(E+\frac{k_1^2}{m n^2}\right)^2+\Gamma^2}-\frac{1}{E^2+\Gamma^2}\Bigg]\Bigg\}, \nonumber
\end{eqnarray}
which exploit the fact that the Breit-Wigner denominator is nearly constant for large $n$. In formula~(\ref{eq:rhofinalform})
the first few perturbative orders are easily identified: since $k_1$ is of order $\as$, the first line contains
the only $0^{\rm th}$ and first order terms, the second one contains the only second and third order terms, and
contains part of the fourth order term. The third line starts with the rest of the fourth order term, since the
expression in the large square bracket is of second order, and the only fifth order term. By expanding the
square bracket one can easily obtain the higher order terms.

Two remarks are in order here. First of all, the formula for $\rho_{\rm cont}$, eq.~(\ref{eq:contfinal}) does not have any
singularity in the limit $\Gamma \to 0$. In fact, in the numerator of each single term in the sum, for negative $E$
and small $\Gamma$ the terms linear in $\Gamma$ cancel, due to the fact that $k_+$ itself becomes of order $\Gamma$
and precisely cancels against the $-|k_1|\Gamma n$ term.

As a second remark, we notice that we cannot interpret the sum in eq.~(\ref{eq:rhofinal}) as the sum over
toponium state. Toponium states are only present in eq.~(\ref{eq:boundfinal}). The combination of the
continuum and bound state contribution is reminiscent of what some of us found in \nrr{}.

\section{Public codes}\label{app:codes}
The code for the three generators that we have introduced are available
in the git repository of the \POWHEGBOX{} distribution (see the
\url{powhegbox.mib.infn.it} for details).
The \thro{} and \thrt{} program are available as targets {\tt pwhg\_main-thr1}
and {\tt pwhg\_main-thr2} of the {\tt Makefile} in the
subdirectory {\tt NonRelativisticCorrections} of the \POWHEGhvq{}
directory.

A subdirectory with the same name is now present also in the
\bbfourl{} process, where
the target that includes the threshold corrections with the highest
accuracy is compiled by default with the name {\tt pwhg\_main}.

Inserting the line {\tt ithreshold <i>} selects the inclusion of threshold
corrections up to and including the $i^{\rm th}$ order in the threshold expansion.
Significant values are 1, 2, 3, 4, 6, and 33, which is the default and represents
the highest accuracy.
The value 34 includes to all orders only the corrections due to the bound states, while
35 includes only the continuum.

For the \thrt{} generator the inclusion of running width effects (dubbed {\tt wc1}
in this work) is active by default. To switch them off one should add the line
{\tt widths\_corr 0} in the {\tt powheg.input} file.

The coupling constant for the threshold corrections is set
automatically in the program as described in the text. It is however
possible to run the program with a fixed value of the coupling used
for the computation of threshold enhanced effects, by adding {\tt
  alphathr <value>} to the powheg.input file. Furthermore, one can
conveniently fix the scale for the coupling by specifying an invariant
mass cut with the input line {\tt alphathrscale <$\mtt^{(\rm max)}$>}, in order to
perform computations of the kind shown in
Sec.~\ref{sec:largeBinsSigma}.

\bibliographystyle{JHEP}
\bibliography{ttbcorr.bib}
 
\end{document}